\documentclass[letterpaper,journal]{IEEEtran}
\usepackage{amsmath,amsfonts,amssymb,bm} 
\usepackage{graphicx} 
\usepackage[caption=false,font=footnotesize]{subfig} 
\usepackage{array,booktabs,multirow} 
\usepackage{cite} 
\usepackage{xcolor} 
\usepackage{amsthm,thmtools} 
\usepackage{algorithm,algorithmic} 
\usepackage{stfloats} 
\usepackage[colorlinks=true,linkcolor=blue,citecolor=blue,urlcolor=blue]{hyperref} 

\declaretheorem[name=Theorem,style=plain]{theorem}
\declaretheorem[name=Lemma,style=plain]{lemma}

\DeclareMathOperator{\blkdiag}{blkdiag}

\begin{document}

\title{Wave Processing Unit: Closed-Form Matrix Synthesis for Wave Computing in Multiuser MIMO Communications}

\author{Xiaokun Teng,\IEEEmembership{}
    Yanqing Ren,\IEEEmembership{}
    Weicong Chen,\IEEEmembership{~Member,~IEEE,}
    Wankai Tang,\IEEEmembership{~Member,~IEEE,}\\
    Xiao Li,\IEEEmembership{~Member,~IEEE}
    and Shi Jin,\IEEEmembership{~Fellow,~IEEE}
    
    
    \thanks{Xiaokun Teng, Yanqing Ren, Weicong Chen, Wankai Tang, Xiao Li and Shi Jin are with the National Mobile Communications Research Laboratory, Southeast University, Nanjing 210096, China (e-mail: xkteng@seu.edu.cn; renyq@seu.edu.cn; cwc@seu.edu.cn; tangwk@seu.edu.cn; li\_xiao@seu.edu.cn; jinshi@seu.edu.cn). \textit{(Corresponding authors: Shi Jin; Wankai Tang.)}}
    
}

\markboth{IEEE Transactions Draft, 2026}{Programmable Planar Mach--Zehnder Interferometer Networks for Wireless Physical-Layer Operator Acceleration}

\maketitle
\begin{abstract}
    Wave computing offers a compelling route to bypass the latency and energy bottlenecks of digital baseband processing. However, existing wave-domain processors rely on task-specific, non-convex iterative optimization, which incurs substantial configuration overhead and limits real-time applications. In this paper, we propose the wave processing unit (WPU), a programmable planar Mach--Zehnder interferometer (MZI) network that accelerates matrix--vector multiplications in MIMO communications. First, we model the tunable MZI element as a parameterized $2\times2$ unitary transformation, and establish a closed-form synthesis framework for the WPU that decomposes an arbitrary unitary matrix into a cascade of MZI elements arranged in a compact rectangular topology. To handle arbitrary complex-valued operators, we further extend the method through singular value decomposition (SVD) and a hardware-efficient semi-unitary synthesis, avoiding redundant hardware and reducing the compilation complexity. We then apply the WPU to multiuser downlink precoding by an efficient WPU synthesis method, and compare the proposed architecture with hybrid and wave-domain ones. Numerical results demonstrate machine-precision synthesis with negligible residuals for up to $64\times64$ operators, with approximately $380\times$ faster configuration than iterative wave computing baselines. For 64-antenna, 8-user downlink precoding, WPU requires 87.5\% fewer RF chains than the fully digital architecture while achieving the same sum rate under ideal hardware. These results highlight the potential of the WPU for precise, hardware-efficient wave-domain processing with rapid configuration.
\end{abstract}


\section{Introduction}
\label{sec:introduction}

As fifth-generation (5G) cellular networks continue to evolve and research advances toward International Mobile Telecommunications for 2030 and beyond (IMT-2030), wireless systems are expected to support sustained traffic growth and an expanding range of data-intensive services. Future networks are envisioned to enable immersive communications, ubiquitous connectivity, integrated sensing and communication, and artificial-intelligence (AI)-enabled services, while meeting increasingly stringent requirements on latency, coverage, and energy efficiency \cite{ericssonMobilityReport2025,iturM2160,tataria2021,letaief2019}. Meeting these requirements calls for wider transmission bandwidths, higher spatial multiplexing orders, and optimized hardware and algorithm co-design for next-generation wireless transceivers. At the physical layer, many signal processing tasks, including precoding, signal detection, and channel estimation, involve matrix operations. Many of these must be repeatedly executed over high-rate symbol streams or across a large number of antennas and subcarriers. As the antenna dimension, processed bandwidth, and number of data streams increase, their digital implementation becomes progressively more challenging in terms of computational complexity, processing latency, and power consumption.

Toward reducing the cost of high-dimensional signal processing, a well-known paradigm has been to move partial spatial processing from the digital baseband to the analog radio-frequency (RF) domain. Classical Butler matrices and microwave lenses implement structured beam-space transformations through passive RF circuits \cite{butler1961,rotman1963}. Lens antenna arrays further exploit electromagnetic (EM) focusing to support spatial multiplexing in multiple-input multiple-output (MIMO) systems \cite{zeng2014lens}. For massive MIMO systems, hybrid analog--digital beamforming combines a low-dimensional digital precoder with an RF phase-shifter network, substantially reducing the number of RF chains and the associated hardware cost \cite{elayach2014spatial,alkhateeb2015limited,gao2016hybrid,molisch2017hybrid}. These architectures have established the value of RF signal processing. 

The evolution of analog processing is now extending from fixed RF networks to reconfigurable EM structures. Instead of treating the antenna response as an immutable part of the transceiver, reconfigurable intelligent surfaces (RISs), fluid antenna systems (FASs), and movable antennas (MAs) \cite{tang2022pathloss, chen2023channel, deng2021holographic, new2025fas,wong2026pixel, zhu2024movable,shao2025sixdma} provide software-controllable EM degrees of freedom (DoF) through position, orientation, or radiation-pattern reconfiguration. The tri-hybrid MIMO architecture further systematizes this direction by complementing conventional digital and analog precoding with an EM-domain processing layer implemented using reconfigurable antennas \cite{castellanos2026trihybrid,heath2026trihybrid}. 

Beyond the above advances, a more ambitious direction is to perform programmable signal processing directly in the wave domain \cite{dardari2026ota}. In this line of research, the desired input--output mapping is embedded into wave propagation, diffraction, or interference, so that the configured EM structure directly participates in the computation. Diffractive deep neural networks (D$^2$NNs) first demonstrated this principle in optical systems by implementing task-oriented transformations through cascaded diffractive layers \cite{lin2018diffractive}. Motivated by this paradigm, stacked intelligent metasurfaces (SIMs) have introduced wave-domain processing into wireless communications, where multiple programmable metasurface layers jointly shape the propagating EM field for holographic MIMO transmission and multiuser beamforming \cite{an2023jsac,an2025twc}. However, stacked three-dimensional metasurface structures also bring practical challenges, including layer alignment, near-field coupling modeling, and integration. Planar diffractive neural networks (PDNNs) address these issues by transferring diffractive wave processing to guided-wave planar circuits, in which engineered couplers and programmable phase shifts implement configurable wave interactions with improved integrability \cite{gu2024pdnn,teng2025pdnn}. Moreover, microwave linear analog computers (MiLACs) show that tunable multiport microwave networks can perform linear-algebraic operations such as matrix inversion, linear estimation, and MIMO beamforming \cite{nerini2025analog,nerini2025analog2}. Recently, programmable multiport microwave circuits have experimentally demonstrated unitary transformation implementation across the 1.5--3.0~GHz band \cite{keshavarz2025programmable}, further verifying the physical feasibility of guided-wave matrix processors. These developments collectively indicate the growing interest of wave-domain analog computation for offloading demanding digital operations, but most existing designs remain task-oriented and optimization-driven.

Despite these advances, a gap remains between task-specific wave-domain processing and universal matrix-operator acceleration in wireless physical-layer systems. Most existing wave-based processors are optimized for a specific communication task. Their circuit parameters are therefore tightly coupled with the chosen objective and are usually obtained through alternating optimization, gradient-based training, or other iterative procedures. Such a design philosophy becomes less efficient when a physical-layer module is required to support a large number of different operators under fast-varying channel conditions. From this perspective, two fundamental limitations emerge. First, existing hardware architectures lack sufficient universality, i.e., the ability to represent an arbitrary matrix rather than only a task-induced transformation. Second, existing approaches typically rely on iterative algorithms for solving a nonconvex optimization problem, which may suffer from convergence uncertainty and substantial computational overhead, negating the latency advantage of wave-domain processing. These limitations highlight the need for wave-domain processing architectures that provide greater representational capacity and efficient synthesis algorithms.

Motivated by the above considerations, this paper introduces the \emph{wave processing unit} (WPU), defined as a reconfigurable microstrip or other guided-wave circuit capable of directly performing mathematical operations on RF signals propagating in waveguides. In particular, we develop a programmable planar Mach--Zehnder interferometer (MZI) network implementation of the WPU. Inspired by universal multiport interferometers \cite{reck1994,clements2016, miller2012, shen2017nanophotonic}, the WPU can realize arbitrary linear transformations through light-speed wave propagation and interference. Compared with other wave-domain processing paradigms, the proposed WPU admits a closed-form synthesis process from a target matrix operator to implementable hardware parameters.

The main contributions of this paper are summarized as follows.

\begin{itemize}
    \item \textbf{Programmable WPU architecture and unitary matrix synthesis:}
    Exploiting wave interference across planar waveguides, we introduce the WPU as a programmable wave computer. We model the transfer matrix of the basic component of the WPU, a tunable symmetric MZI element, and establish a closed-form synthesis framework from an arbitrary unitary matrix to a compact WPU. We derive the conditions for sequential matrix elimination and MZI standardization that enable exact synthesis.
    
    \item \textbf{Efficient closed-form synthesis of general complex matrices:} We extend the closed-form synthesis framework to arbitrary non-unitary matrices through SVD and semi-unitary synthesis. For an $M \times N$ matrix operator with $M < N$, the proposed semi-unitary synthesis avoids completing the larger-side singular factor into a redundant $N \times N$ unitary matrix. It saves ${(N-M)(N-M-1)}/{2}$ programmable MZI elements and reduces the corresponding compilation complexity to $\mathcal{O}(NM^2)$.
    
    \item \textbf{Structure-aware WPU synthesis for multiuser MMSE precoding:}
    We specialize the general framework to multiuser downlink precoding by exploiting the structural characteristic of the typical MMSE precoding matrix. Rather than explicitly constructing the MMSE precoder and decomposing it for WPU synthesis, we derive the WPU configuration directly from the compact SVD of the downlink channel, thereby avoiding matrix inversion calculation of MMSE precoder. Compared with other hybrid or wave-based precoding architectures, the proposed WPU implements full DoFs with the minimal RF-chain and closed-form configuration.
    
    \item \textbf{Numerical evaluation and comparison with other wave-based architectures:}
    We benchmark the WPU against SIM, PDNN, hybrid beamforming, and fully digital architectures across various scales. At the operator level, closed-form synthesis achieves machine-precision accuracy with approximately $380\times$ faster configuration than iterative wave computing baselines for $64\times 64$ matrices, while semi-unitary synthesis reduces MZI element count by 52.4\% for $16\times64$ operators. At the link level, for a 64-antenna, 8-user downlink precoding, the WPU eliminates 87.5\% of the RF chains required by fully digital precoding while sustaining identical sum rates. We further evaluate the influence of potential hardware-induced perturbations and show that hardware errors primarily cause an interference-limited rate ceiling.
\end{itemize}

The remainder of this paper is organized as follows. Section~\ref{sec:architecture} introduces the planar MZI architecture and develops the closed-form synthesis procedure for arbitrary unitary operators. Section~\ref{sec:non_unitary_synthesis} extends this framework to semi-unitary and arbitrary complex-valued matrix operators. Section~\ref{sec:precoding} applies the WPU to multiuser downlink precoding and evaluates the corresponding communication performance and latency characteristics. Section~\ref{sec:numerical_results} presents numerical results and comparisons with optimization-based SIM and PDNN architectures. Finally, Section~\ref{sec:conclusion} concludes the paper.

\section{Architecture and Unitary Matrix Synthesis of the Wave Processing Unit}
\label{sec:architecture}
The WPU accelerates matrix--vector multiplications (MVMs) by shifting them from the digital baseband to the EM wave domain. To achieve this, the hardware architecture should represent prescribed matrix operators through a deterministic hardware layout. This section establishes the mapping from arbitrary unitary operators to the circuit topology and parameters of the WPU.

\begin{figure*}[!t]
    \centering
    \includegraphics[width=0.9\textwidth]{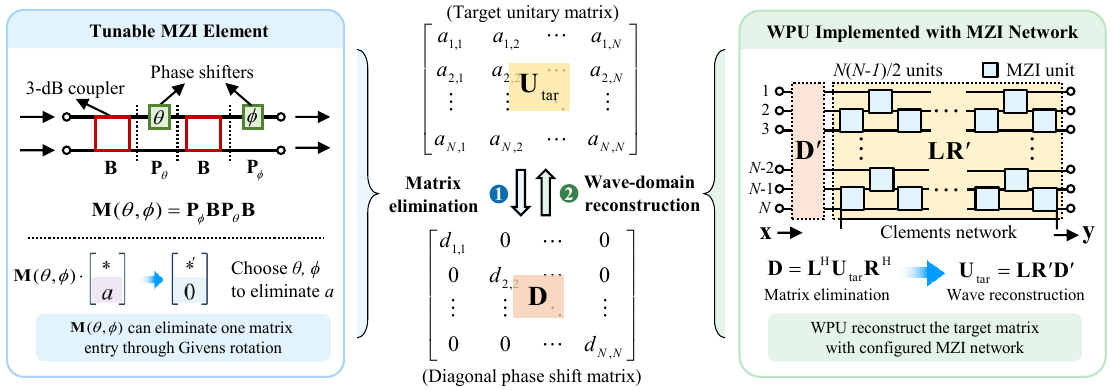}
    \caption{The principle of WPU-based unitary transformation. A tunable MZI element, comprising 2 couplers and 2 phase shifters, implements an unitary transformation between two adjacent signal paths. Cascading such elements in a Clements network reduces the target unitary matrix \(\mathbf{U}_{\text{tar}}\) to the diagonal phase shift matrix \(\mathbf{D}\) during synthesis procedure. Programming the resulting MZI parameters reconstructs \(\mathbf{U}_{\text{tar}}\) for the wave-domain MVM \(\mathbf{y}=\mathbf{U}_{\text{tar}}\mathbf{x}\).}
    \label{fig:primaries_overview}
\end{figure*}

\subsection{Tunable MZI Element}
\label{subsec:MZIunit}
The elementary circuit of the WPU is the tunable symmetric MZI shown in Fig.~\ref{fig:primaries_overview}. It consists of two 3-dB directional couplers and two phase shifters. Let \(\mathbf{B}\) denote the narrow-band transfer matrix of an ideal 3-dB coupler, let \(\mathbf{P}_{\theta}\) denote the internal phase-shift matrix, and let \(\mathbf{P}_{\phi}\) denote the external phase-shift matrix, i.e.,
\begin{equation}
    \label{eq:coupler_revised}
    \mathbf{B} = \frac{1}{\sqrt{2}}
    \begin{bmatrix}
        1 & j\\
        j & 1
    \end{bmatrix},
    \, \mathbf{P}_{\theta} =
    \begin{bmatrix}
        e^{j\theta} & 0\\
        0 & 1
    \end{bmatrix},
    \, \mathbf{P}_{\phi} =
    \begin{bmatrix}
        e^{j\phi} & 0\\
        0 & 1
    \end{bmatrix}.
\end{equation}
Here, \(\theta\in[0,\pi]\) denotes the internal phase shift, whereas \(\phi\in[0,2\pi)\) denotes the external phase shift, $j=\sqrt{-1}$. The transfer matrix of the MZI element is therefore
\begin{equation}
    \label{eq:mzi_block_revised}
    \mathbf{M}(\theta,\phi) = \mathbf{P}_{\phi}\mathbf{B}\mathbf{P}_{\theta}\mathbf{B},
\end{equation}
which can be written explicitly as
\begin{equation}
    \label{eq:mzi_block_explicit_revised}
    \mathbf{M}(\theta,\phi) = j e^{j\theta/2}
    \begin{bmatrix}
        e^{j\phi}\sin(\theta/2) & e^{j\phi}\cos(\theta/2)\\
        \cos(\theta/2) & -\sin(\theta/2)
    \end{bmatrix}.
\end{equation}
Since \(\mathbf{B}\), \(\mathbf{P}_{\theta}\), and \(\mathbf{P}_{\phi}\) are all unitary, \(\mathbf{M}(\theta,\phi)\) is unitary for every \((\theta,\phi)\). By adjusting the phase shifts $\theta$ and $\phi$, the MZI element controls the wave interference between two adjacent signal paths. More importantly, this transfer matrix in \eqref{eq:mzi_block_explicit_revised} can be parameterized to annihilate an arbitrary element within a target matrix by applying Givens rotation \cite{golub2013matrix}. In the following subsection, we explain how this principle is used to construct an MZI network capable of implementing any unitary matrix.

\subsection{Unitary Matrix Synthesis via Rectangular MZI Network}

By cascading the MZI elements described in Section \ref{subsec:MZIunit}, we can build a universal planar MZI network for any unitary MVM. Related structures include the triangular network proposed by Reck \textit{et al.} \cite{reck1994} and the rectangular network proposed by Clements \textit{et al.} \cite{clements2016}, which is shown in Fig.~\ref{fig:primaries_overview}. They use the same number of $N(N-1)/2$ tunable MZI elements to implement an $N\times N$ unitary matrix, but they arrange these elements in different shapes. In the rectangular network, the longest propagation path crosses only \(N\) elements instead of \(2N-3\) elements in the triangular one for realizing an \(N\times N\) matrix \cite{clements2016}. This reduces insertion-loss accumulation and yields a more compact layout. For this reason, the present paper adopts the Clements rectangular layout as the architecture for unitary-matrix synthesis.

Algorithm~\ref{alg:clements_revised} summarize the unitary matrix synthesis procedure based on the rectangular MZI network, which starts with a sequence of Givens rotations applied to the target unitary matrix $\mathbf{U}_{\text{tar}}$. To describe the synthesis algorithm, the MZI transfer matrix in \eqref{eq:mzi_block_explicit_revised} is embedded into the \(N\times N\) network as
\begin{equation}
    \label{eq:embed_adjacent_revised}
    \mathbf{G}_{p}(\theta,\phi) = \blkdiag\!\big( \mathbf{I}_{p-1}, \mathbf{M}(\theta,\phi), \mathbf{I}_{N-p-1} \big) \in \mathbb{C}^{N\times N},
\end{equation}
for \(p=1,\ldots,N-1\), where \(\mathbf{G}_{p}(\theta,\phi)\) acts on dimensions \(p\) and \(p+1\) only. 

Let $\mathbf{W} \in \mathbb{C}^{N \times N}$ denote the working matrix, initialized as $\mathbf{W}=\mathbf{U}_{\text{tar}}$. The following lemma demonstrates how the embedded matrix acts on adjacent rows or columns to achieve matrix elimination, where we define $\angle 0 = 0$.

\begin{lemma}
    \label{thm:row_annihilation}
    Suppose that the entry \([\mathbf{W}]_{r,c}\) is to be eliminated by mixing rows \(r-1\) and \(r\), where \(2\le r\le N\). By choosing
    \begin{equation}
        \label{eq:left_params_revised}
        \phi_{r,c}^{\mathrm{row}} = \operatorname{mod} (\angle[\mathbf{W}]_{r-1,c} - \angle[\mathbf{W}]_{r,c}, 2\pi),
    \end{equation}
    \begin{equation}
        \label{eq:left_params_angle_revised}
        \theta_{r,c}^{\mathrm{row}} = 2\arctan \frac{|[\mathbf{W}]_{r-1,c}|} {|[\mathbf{W}]_{r,c}|},
    \end{equation}
    (\(\theta_{r,c}^{\mathrm{row}}=\pi\) if \([\mathbf{W}]_{r,c}=0\)), then the update
    \begin{equation}
        \label{eq:left_action_revised}
        \mathbf{W}' = \left( \mathbf{G}_{r-1} \big( \theta_{r,c}^{\mathrm{row}}, \phi_{r,c}^{\mathrm{row}} \big) \right)^\mathrm{H} \mathbf{W}
    \end{equation}
    enforces \([\mathbf{W}']_{r,c}=0\).
\end{lemma}
\begin{proof}
    See Appendix~\ref{app:row_annihilation}.
\end{proof}
\begin{lemma}
    \label{thm:col_annihilation}
    Suppose that the entry \([\mathbf{W}]_{r,c}\) is to be eliminated by mixing columns \(c\) and \(c+1\), where \(1\le c\le N-1\). By choosing
    \begin{equation}
        \label{eq:right_params_revised}
        \phi_{r,c}^{\mathrm{col}} = \operatorname{mod} (\angle[\mathbf{W}]_{r,c+1} - \angle[\mathbf{W}]_{r,c} + \pi, 2\pi),
    \end{equation}
    \begin{equation}
        \label{eq:right_params_angle_revised}
        \theta_{r,c}^{\mathrm{col}} = 2\arctan \frac{|[\mathbf{W}]_{r,c+1}|} {|[\mathbf{W}]_{r,c}|},
    \end{equation}
    (\(\theta_{r,c}^{\mathrm{col}}=\pi\) if \([\mathbf{W}]_{r,c}=0\)), then the update
    \begin{equation}
        \label{eq:right_action_revised}
        \mathbf{W}' = \mathbf{W} \mathbf{G}_{c} \big( \theta_{r,c}^{\mathrm{col}}, \phi_{r,c}^{\mathrm{col}} \big)
    \end{equation}
    enforces \([\mathbf{W}']_{r,c}=0\).
\end{lemma}
\begin{proof}
    See Appendix~\ref{app:col_annihilation}.
\end{proof}

Applying Lemmas~\ref{thm:row_annihilation} and~\ref{thm:col_annihilation} in the order specified by the Step~1 of Algorithm~\ref{alg:clements_revised} enables the annihilation of all off-diagonal elements in the lower-triangular portion of the target matrix. Starting from the lower-left corner, the $s$-th iteration nullifies $s$ entries along the line parallel to the main diagonal, namely those indexed by $(N-s+p, p)$ for $p = 1, \dots, s$. After $N-1$ steps, the algorithm nullifies all off-diagonal lower-triangular entries of $\mathbf{U}_{\text{tar}}$. This elimination order ensures that values of the previously eliminated entries are not reactivated in subsequent steps. Meanwhile, the alternation between adjacent-row and adjacent-column elimination is introduced to form the rectangular hardware layout in Fig.~\ref{fig:primaries_overview}. 

A key observation is that the working matrix remains unitary throughout the elimination. Indeed, \(\mathbf{G}_{p}(\theta,\phi)\) is unitary because it is obtained by embedding the unitary \(2\times2\) block \(\mathbf{M}(\theta,\phi)\) into a larger identity matrix. Therefore, every update in Algorithm~\ref{alg:clements_revised} multiplies the current working matrix by a unitary matrix, so if \(\mathbf{W}=\mathbf{U}_{\text{tar}}\) is unitary at initialization, then it remains unitary after every update.

After all \(N-1\) stages are completed, the final working matrix \(\mathbf{W}\) is upper triangular. Since the elimination process preserves unitarity, \(\mathbf{W}\) is in fact an upper triangular unitary matrix. Such a matrix must be diagonal. Hence
\begin{equation}
    \label{eq:final_diagonal_revised}
    \mathbf{D} = \mathbf{L}^\mathrm{H}\mathbf{U}_{\text{tar}}\mathbf{R}^\mathrm{H},
\end{equation}
is also diagonal, where $\mathbf{L}^\mathrm{H}$ and $\mathbf{R}^\mathrm{H}$ collect the unitary matrices generated by adjacent-row and adjacent-column eliminations, respectively. Although the algorithm only explicitly eliminates the lower-triangular off-diagonal entries, the upper-triangular ones disappear implicitly because the final matrix remains unitary.

\begin{algorithm}[!t]
    \caption{Closed-Form Unitary Synthesis on Clements Rectangular Network}
    \label{alg:clements_revised}
    \begin{algorithmic}[1]
        \STATE \textbf{Input:} Target unitary matrix \(\mathbf{U}_{\text{tar}}\in\mathbb{C}^{N\times N}\)
        
        \STATE \textbf{// Step 1: Alternating Row/Column Elimination}
        \STATE Initialize working matrix \(\mathbf{W}\leftarrow \mathbf{U}_{\text{tar}}\)
        \FOR{\(s=1\) to \(N-1\)}
        \IF{\(s\) is odd}
        \FOR{\(p=1\) to \(s\)}
        \STATE Set \(c\leftarrow p\) and \(r\leftarrow N-s+p\)
        \STATE Compute \(\big(\theta_{r,c}^{\mathrm{row}},\phi_{r,c}^{\mathrm{row}}\big)\) from \eqref{eq:left_params_revised} and \eqref{eq:left_params_angle_revised}
        \STATE Update \(\mathbf{W}\leftarrow \big(\mathbf{G}_{r-1}(\theta_{r,c}^{\mathrm{row}},\phi_{r,c}^{\mathrm{row}})\big)^\mathrm{H}\mathbf{W}\)
        \ENDFOR
        \ELSE
        \FOR{\(p=s\) down to \(1\)}
        \STATE Set \(c\leftarrow p\) and \(r\leftarrow N-s+p\)
        \STATE Compute \(\big(\theta_{r,c}^{\mathrm{col}},\phi_{r,c}^{\mathrm{col}}\big)\) from \eqref{eq:right_params_revised} and \eqref{eq:right_params_angle_revised}
        \STATE Update \(\mathbf{W} \leftarrow \mathbf{W} \mathbf{G}_{c}(\theta_{r,c}^{\mathrm{col}},\phi_{r,c}^{\mathrm{col}})\)
        \ENDFOR
        \ENDIF
        \ENDFOR
        
        \STATE Set \(\mathbf{D}\leftarrow \mathbf{W}\)
        \STATE Form \(\mathbf{L}\) and \(\mathbf{R}\) from \eqref{eq:L_full} and \eqref{eq:R_full}
        
        \STATE \textbf{// Step 2: MZI Element Standardization}
        \STATE Initialize running diagonal phase matrix \(\mathbf{D}'\leftarrow \mathbf{D}\)
        \FOR{\(s=N-1\) down to \(1\)}
        \IF{\(s\) is even}
        \FOR{\(p=1\) to \(s\)}
        \STATE Compute \(\phi_{N-s+p,p}^{\prime\mathrm{col}} \) according to \eqref{eq:embedded_push_phi}
        \STATE Update \(\mathbf{D}'\) according to \eqref{eq:embedded_push_d1}--\eqref{eq:embedded_push_d2}
        \ENDFOR
        \ENDIF
        \ENDFOR
        \STATE Form standardized \(\mathbf{R}'\) via \eqref{eq:R_prime_full}
        
        \STATE \textbf{Output:} All MZI phase parameters and \((\mathbf{L},\mathbf{R}',\mathbf{D}')\)
    \end{algorithmic}
\end{algorithm}

Reconstructing the target matrix with all the elimination matrices yields
\begin{equation}
    \label{eq:clements_factorization_revised}
    \mathbf{U}_{\text{tar}} = \mathbf{L}\mathbf{D}\mathbf{R}.
\end{equation}
By defining $\prod_{k=a}^{b}\mathbf{X}_k:=\mathbf{X}_a\cdots\mathbf{X}_b$, $\mathbf{L}$ and $\mathbf{R}$ can be expanded as 
\begin{equation}
    \label{eq:L_full}
    \mathbf{L} = \prod_{\substack{s=1\\ s\ \mathrm{odd}}}^{N-1}
    \left( \prod_{p=1}^{s} \mathbf{G}_{N-s+p-1} \big( \theta_{N-s+p,p}^{\mathrm{row}}, \phi_{N-s+p,p}^{\mathrm{row}} \big) \right),
\end{equation}
and
\begin{equation}
    \label{eq:R_full}
    \mathbf{R} = \prod_{\substack{s=N-1\\ s\ \mathrm{even}}}^{1}
    \left( \prod_{p=1}^{s} \bigg( \mathbf{G}_{p} \big( \theta_{N-s+p,p}^{\mathrm{col}}, \phi_{N-s+p,p}^{\mathrm{col}} \big) \bigg)^\mathrm{H} \right).
\end{equation}
Here, \(\mathbf{L}\) is already expressed in the standard hardware form \(\mathbf{G}_{p}(\theta,\phi)\). By contrast, \(\mathbf{R}\) still contains conjugate-transposed embedded factors. Consequently, one standardization step is needed before both $\mathbf{L}$ and $\mathbf{R}$ can be implemented by the standard MZI element. This conversion is established next by propagating the diagonal phase layer $\mathbf{D}$ through the factors in \(\mathbf{R}\).
\begin{lemma}
    \label{thm:embedded_push}
    Let \(\boldsymbol{\Delta} = \operatorname{diag}(d_1,\ldots,d_N)\) be a diagonal unitary matrix. For the embedded MZI transformation \(\mathbf{G}_{p}(\theta,\phi)\) defined in \eqref{eq:embed_adjacent_revised}, there exists \(\boldsymbol{\Delta}' = \operatorname{diag}(d_1,\ldots, d'_p, d'_{p+1} ,\ldots,d_N)\) and an external phase $\phi'$ such that
    \begin{equation}
        \label{eq:embedded_push_rule}
        \boldsymbol{\Delta} \left( \mathbf{G}_{p}(\theta,\phi) \right)^\mathrm{H} = \mathbf{G}_{p}(\theta,\phi') \boldsymbol{\Delta}',
    \end{equation}
    where
    \begin{equation}
        \label{eq:embedded_push_phi}
        \phi' = \operatorname{mod} ( \angle d_p - \angle d_{p+1}, 2\pi ),
    \end{equation}
    \begin{equation}
        \label{eq:embedded_push_d1}
        d_p' = -e^{-j(\theta+\phi)}\,d_{p+1},
    \end{equation}
    \begin{equation}
        \label{eq:embedded_push_d2}
        d_{p+1}' = -e^{-j\theta}\,d_{p+1},
    \end{equation}
    with $d_q$ unchanged, for $q\in\{1,\dots,N\} \backslash \{p,p+1\}$.
\end{lemma}
\begin{proof}
    See Appendix~\ref{app:diagonal_push}.
\end{proof}

Lemma~\ref{thm:embedded_push} implies that whenever the current diagonal matrix appears immediately to the left of a factor \(\big(\mathbf{G}_{p}(\theta,\phi)\big)^\mathrm{H}\) in \(\mathbf{R}\), \eqref{eq:embedded_push_rule} replaces that factor by a standard embedded MZI transformation \(\mathbf{G}_{p}(\theta,\phi')\) and updates two diagonal entries of the current diagonal matrix. In this transformation, the internal phase parameter $\theta$ remains unchanged, while the external one $\phi$ is reassigned.

Applying Lemma~\ref{thm:embedded_push} sequentially to all factors in \(\mathbf{R}\) yields
\begin{equation}
    \label{eq:DR_to_RprimeDprime}
    \mathbf{D}\mathbf{R} = \mathbf{R}'\mathbf{D}',
\end{equation}
where
\begin{equation}
    \label{eq:R_prime_full}
    \mathbf{R}' = \prod_{\substack{s=N-1\\ s\ \mathrm{even}}}^{1}
    \left( \prod_{p=1}^{s} \mathbf{G}_{p} \big( \theta_{N-s+p,p}^{\prime\mathrm{col}}, \phi_{N-s+p,p}^{\prime\mathrm{col}} \big) \right),
\end{equation}
with $\theta_{N-s+p,p}^{\prime\mathrm{col}}=\theta_{N-s+p,p}^{\mathrm{col}}$. Each \(\phi_{N-s+p,p}^{\prime\mathrm{col}}\) is calculated through \eqref{eq:embedded_push_phi}. Substituting \eqref{eq:DR_to_RprimeDprime} into \eqref{eq:clements_factorization_revised} gives the final hardware-consistent factorization
\begin{equation}
    \label{eq:clements_factorization_LRprimeDprime}
    \mathbf{U}_{\text{tar}} = \mathbf{L}\mathbf{R}'\mathbf{D}'.
\end{equation}

Equation~\eqref{eq:clements_factorization_LRprimeDprime} completes the synthesis process described by Algorithm~\ref{alg:clements_revised}. In particular, both \(\mathbf{L}\) and \(\mathbf{R}'\) are now realized by the same tunable MZI element \(\mathbf{M}(\theta,\phi)\), while \(\mathbf{D}'\) is implemented as an array composed of $N$ phase shifters. Accordingly, the algorithm outputs a complete set of implementable device parameters: the left-matrix parameters \(\big(\theta_{r,c}^{\mathrm{row}},\phi_{r,c}^{\mathrm{row}}\big)\), the standardized right-matrix parameters \(\big(\theta_{r,c}^{\prime\mathrm{col}},\phi_{r,c}^{\prime\mathrm{col}}\big)\), and the diagonal phases contained in \(\mathbf{D}'\). The resulting factorization corresponds directly to the Clements layout in Fig.~\ref{fig:primaries_overview}.

\section{Non-Unitary Matrix Synthesis}
\label{sec:non_unitary_synthesis}

Many wireless communication algorithms involve general complex-valued matrix operators rather than only unitary transformations. We therefore extend the synthesis framework in Section~\ref{sec:architecture} to non-unitary matrices. 

\subsection{Synthesis Procedure}
We use the SVD to synthesize an arbitrary nonzero matrix \(\mathbf{A}\in\mathbb{C}^{M\times N}\), with \(\mathbf{A}\ne\mathbf{0}\). For a passive realization, the singular values must not exceed unity. We therefore first normalize \(\mathbf{A}\) by its largest singular value:
\begin{equation}
    \label{eq:contractive_norm_revised}
    \mathbf{A} = \kappa_{\text{A}} \widetilde{\mathbf{A}}, \qquad \kappa_{\text{A}} = \sigma_{\max}(\mathbf{A}).
\end{equation}
In the following, we synthesize \(\widetilde{\mathbf{A}}\) instead; the scalar \(\kappa_{\text{A}}\) can be restored outside the passive MZI network by amplification if needed. Applying SVD to \(\widetilde{\mathbf{A}}\) yields
\begin{equation}
    \label{eq:svd_rectangular_revised}
    \widetilde{\mathbf{A}} = \mathbf{U}_{\text{A}}\boldsymbol{\Sigma}\mathbf{V}_{\text{A}}^\mathrm{H},
\end{equation}
where \(\mathbf{U}_{\text{A}}\in\mathbb{C}^{M\times M}\) and \(\mathbf{V}_{\text{A}}\in\mathbb{C}^{N\times N}\) are unitary.

Depending on the relationship between $M$ and $N$, the synthesis of \eqref{eq:svd_rectangular_revised} naturally falls into three cases:
\begin{itemize}
    \item \textbf{Case 1 (\(M=N\)):} Both \(\mathbf{U}_{\text{A}}\) and \(\mathbf{V}_{\text{A}}\) are \(N\times N\) square unitary matrices, and \(\boldsymbol{\Sigma}=\operatorname{diag}(\sigma_1,\ldots,\sigma_N)\). In this case, \(\mathbf{U}_{\text{A}}\) and \(\mathbf{V}_{\text{A}}^\mathrm{H}\) are directly synthesized using Algorithm~\ref{alg:clements_revised}, while \(\boldsymbol{\Sigma}\) is realized by \(N\) parallel single-channel scalers (detailed later in this subsection).
    \item \textbf{Case 2 (\(M<N\)):} The singular value matrix is given by \(\boldsymbol{\Sigma}=\big[\bar{\boldsymbol{\Sigma}}\ \mathbf{0}_{M\times(N-M)}\big]\), where \(\bar{\boldsymbol{\Sigma}}=\operatorname{diag}(\sigma_1,\ldots,\sigma_M)\) with \(0\le \sigma_i\le 1\). Only the smaller-dimensional factor \(\mathbf{U}_{\text{A}}\) requires a full \(M\times M\) unitary mesh, whereas the larger-dimensional factor \(\mathbf{V}_{\text{A}}\) can be partitioned to enable a hardware-efficient semi-unitary realization, especially when \(M\ll N\).
    \item \textbf{Case 3 (\(M>N\)):} The singular value matrix is given by \(\boldsymbol{\Sigma}=\big[\bar{\boldsymbol{\Sigma}}\ \mathbf{0}_{N\times(M-N)}\big]^\mathrm{T}\), where \(\bar{\boldsymbol{\Sigma}}=\operatorname{diag}(\sigma_1,\ldots,\sigma_N)\) with \(0\le \sigma_i\le 1\). Symmetrically, only the smaller-dimensional factor \(\mathbf{V}_{\text{A}}\) requires a full \(N\times N\) unitary mesh, whereas the larger-dimensional factor \(\mathbf{U}_{\text{A}}\) can be partitioned into an \(M\times N\) semi-unitary factor for hardware-efficient realization.
\end{itemize}

We first focus on the non-trivial rectangular case with \(M < N\). Partition \(\mathbf{V}_{\text{A}} = [\mathbf{V}_{\text{A},1}\ \mathbf{V}_{\text{A},2}]\), where \(\mathbf{V}_{\text{A},1}\in\mathbb{C}^{N\times M}\). Then \eqref{eq:svd_rectangular_revised} becomes
\begin{equation}
    \label{eq:thin_svd_revised}
    \widetilde{\mathbf{A}} = \mathbf{U}_{\text{A}}\bar{\boldsymbol{\Sigma}}\mathbf{V}_{\text{A},1}^\mathrm{H}.
\end{equation}
Moreover, because \(\mathbf{V}_{\text{A}}\) is unitary, the column block \(\mathbf{V}_{\text{A},1}\) satisfies \(\mathbf{V}_{\text{A},1}^\mathrm{H}\mathbf{V}_{\text{A},1} = \mathbf{I}_{M}\) and is therefore semi-unitary. Hence, only the left factor \(\mathbf{U}_{\text{A}}\) needs a full \(M\times M\) unitary realization, whereas the right factor \(\mathbf{V}_{\text{A},1}^\mathrm{H}\) only requires a semi-unitary realization. The overall compilation pipeline for synthesizing a rectangular matrix $\widetilde{\mathbf{A}}$ with $M<N$ is outlined in Algorithm~\ref{alg:proposed_arbitrary_revised}.

To describe the decomposition, define the embedding matrix
\begin{equation}
    \label{eq:rectangular_embedding_revised}
    \mathbf{E}_{N,M} =
    \begin{bmatrix}
        \mathbf{I}_{M}\\
        \mathbf{0}_{(N-M)\times M}
    \end{bmatrix}
    \in\mathbb{C}^{N\times M}.
\end{equation}
The key step is to synthesize an arbitrary semi-unitary matrix \(\mathbf{Q}\in\mathbb{C}^{N\times M}\) with \(\mathbf{Q}^\mathrm{H}\mathbf{Q}=\mathbf{I}_{M}\) directly, without first completing it into a full \(N\times N\) unitary matrix.

\begin{theorem}
    \label{thm:semi_unitary_synthesis}
    Let \(\mathbf{Q}\in\mathbb{C}^{N\times M}\) satisfy \(\mathbf{Q}^\mathrm{H}\mathbf{Q}=\mathbf{I}_{M}\) with \(N > M\). Then there exist a diagonal unitary matrix \(\mathbf{D}_{\mathrm Q} = \operatorname{diag}(d_{1},\ldots,d_{M}) \in\mathbb{C}^{M\times M}\) and a product of embedded MZI transformations
    \begin{equation}
        \label{eq:LQ_full}
        \mathbf{L}_{\mathrm Q} = \prod_{c=1}^{M} \left( \prod_{r=N}^{c+1} \mathbf{G}_{r-1} \big( \theta_{r,c}^{\mathrm Q}, \phi_{r,c}^{\mathrm Q} \big) \right),
    \end{equation}
    such that
    \begin{equation}
        \label{eq:semi_factorization_revised}
        \mathbf{Q} = \mathbf{L}_{\mathrm Q} \mathbf{E}_{N,M} \mathbf{D}_{\mathrm Q}.
    \end{equation}
    Here, every pair \((\theta_{r,c}^{\mathrm Q},\phi_{r,c}^{\mathrm Q})\) is obtained in closed form by applying Lemma~\ref{thm:row_annihilation} to the current \(N\times M\) working matrix.
\end{theorem}
\begin{proof}
    See Appendix~\ref{app:semi_unitary_synthesis}.
\end{proof}

Unlike the unitary synthesis in Section~\ref{sec:architecture}, which alternates between adjacent-row and adjacent-column eliminations to form a rectangular mesh, Theorem~\ref{thm:semi_unitary_synthesis} relies solely on one-sided adjacent-row operations. Specifically, this procedure eliminates the sub-diagonal entries within the active \(N\times M\) columns, transforming the semi-unitary matrix directly into the embedded diagonal form \(\mathbf{E}_{N,M}\mathbf{D}_{\mathrm Q}\). The step-by-step semi-unitary elimination routine is given in Step~3 of Algorithm~\ref{alg:proposed_arbitrary_revised}.

\begin{figure*}[!t]
    \centering
    \includegraphics[width=0.9\textwidth]{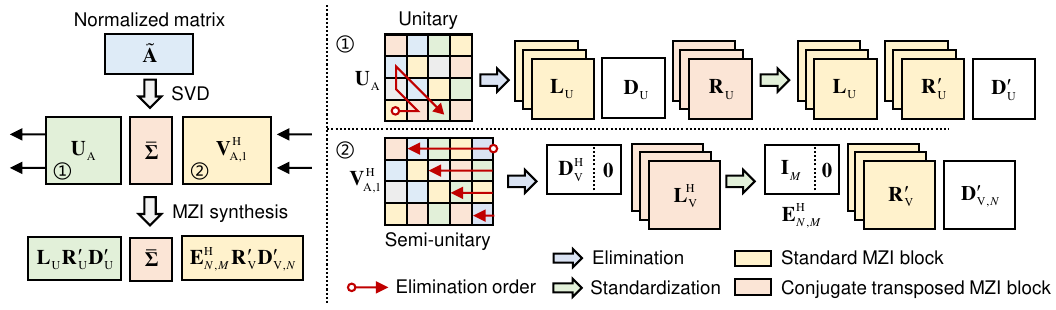}
    \caption{Closed-form synthesis of a normalized non-unitary operator \(\widetilde{\mathbf{A}}\in\mathbb{C}^{M\times N}\) with \(M<N\). The left panel shows its SVD into the square unitary factor \(\mathbf{U}_{\text A}\), the diagonal singular-value layer \(\bar{\boldsymbol{\Sigma}}\), and the semi-unitary factor \(\mathbf{V}_{\text A,1}^{\mathrm H}\), together with their hardware-consistent factors. The right panel illustrates the elimination and standardization procedures that compile \(\mathbf{U}_{\text A}\) into \(\mathbf{L}_{\mathrm U}\mathbf{R}_{\mathrm U}'\mathbf{D}_{\mathrm U}'\) and \(\mathbf{V}_{\text A,1}^{\mathrm H}\) into \(\mathbf{E}_{N,M}^{\mathrm H}\mathbf{R}_{\mathrm V}'\mathbf{D}_{\mathrm V,N}'\).}
    \label{fig:non_unitary_synthesis}
\end{figure*}

Combining the standard unitary synthesis in Section~\ref{sec:architecture} with the above semi-unitary synthesis establishes the complete hardware realization for the non-unitary matrix $\mathbf{A}$ with $M<N$. Applying \eqref{eq:clements_factorization_LRprimeDprime} to the square unitary factor \(\mathbf{U}_{\text{A}}\) in \eqref{eq:thin_svd_revised}, we obtain
\begin{equation}
    \label{eq:U_factor_revised}
    \mathbf{U}_{\text{A}} = \mathbf{L}_{\mathrm U} \mathbf{R}_{\mathrm U}' \mathbf{D}_{\mathrm U}',
\end{equation}
where \(\mathbf{L}_{\mathrm U}\) and \(\mathbf{R}_{\mathrm U}'\) are standard \(M\times M\) Clements meshes and \(\mathbf{D}_{\mathrm U}'\) is a diagonal phase layer. Meanwhile, applying \eqref{eq:semi_factorization_revised} to \(\mathbf{V}_{\text{A},1}\) and denoting the resulting factors by \(\mathbf{L}_{\mathrm V}\) and \(\mathbf{D}_{\mathrm V}\), its Hermitian transpose is
\begin{equation}
    \mathbf{V}_{\text{A},1}^\mathrm{H} = \mathbf{D}_{\mathrm V}^\mathrm{H} \mathbf{E}_{N,M}^\mathrm{H} \mathbf{L}_{\mathrm V}^\mathrm{H}.
\end{equation}
To standardize the conjugate-transposed MZI factors in \(\mathbf{L}_{\mathrm V}^\mathrm{H}\), we extend \(\mathbf{D}_{\mathrm V}\) to the \(N\times N\) diagonal matrix
\begin{equation}
    \label{eq:Ds_extension_revised}
    \mathbf{D}_{\mathrm V,N} = \operatorname{diag}(d_{1},\ldots,d_{M},1,\ldots,1).
\end{equation}
Then, \(\mathbf{V}_{\text{A},1}^\mathrm{H} = \mathbf{E}_{N,M}^\mathrm{H} \mathbf{D}_{\mathrm V,N}^\mathrm{H} \mathbf{L}_{\mathrm V}^\mathrm{H}\). Applying Lemma~\ref{thm:embedded_push} sequentially to \(\mathbf{D}_{\mathrm V,N}^\mathrm{H}\mathbf{L}_{\mathrm V}^\mathrm{H}\) yields
\begin{equation}
    \label{eq:semi_standardization_revised}
    \mathbf{D}_{\mathrm V,N}^\mathrm{H} \mathbf{L}_{\mathrm V}^\mathrm{H} = \mathbf{R}_{\mathrm V}' \mathbf{D}_{\mathrm V,N}',
\end{equation}
where \(\mathbf{D}_{\mathrm V,N}'\) is diagonal and \(\mathbf{R}_{\mathrm V}'\) is a product of standard embedded MZI transformations given by
\begin{equation}
    \label{eq:Rs_prime_full_revised}
    \mathbf{R}_{\mathrm V}' = \prod_{c=M}^{1} \left( \prod_{r=c+1}^{N} \mathbf{G}_{r-1} \big( \theta_{r,c}^{\mathrm V}, \phi_{r,c}^{\prime\mathrm V} \big) \right).
\end{equation}
Therefore, we have
\begin{equation}
    \label{eq:V1H_standardized_revised}
    \mathbf{V}_{\text{A},1}^\mathrm{H} = \mathbf{E}_{N,M}^\mathrm{H} \mathbf{R}_{\mathrm V}' \mathbf{D}_{\mathrm V,N}'.
\end{equation}
Substituting \eqref{eq:U_factor_revised} and \eqref{eq:V1H_standardized_revised} into \eqref{eq:thin_svd_revised} and defining \(\boldsymbol{\Gamma} = \mathbf{D}_{\mathrm U}'\bar{\boldsymbol{\Sigma}}\), we arrive at the final hardware-consistent factorization
\begin{equation}
    \label{eq:proposed_arbitrary_factorization_scaled_revised}
    \mathbf{A} = \kappa_{\text{A}} \mathbf{L}_{\mathrm U} \mathbf{R}_{\mathrm U}' \boldsymbol{\Gamma} \mathbf{E}_{N,M}^\mathrm{H} \mathbf{R}_{\mathrm V}' \mathbf{D}_{\mathrm V,N}'.
\end{equation}
The complete factorization and the corresponding unitary and semi-unitary synthesis paths are summarized in Fig.~\ref{fig:non_unitary_synthesis}.

\begin{algorithm}[!t]
    \caption{Closed-Form Synthesis for Non-Square Matrices}
    \label{alg:proposed_arbitrary_revised}
    \begin{algorithmic}[1]
        \STATE \textbf{Input:} \(\mathbf{A}\in\mathbb{C}^{M\times N}\), \(\mathbf{A}\ne\mathbf{0}\), with \(M < N\)
        \STATE \textbf{// Step 1: Pre-scaling and SVD}
        \STATE Compute \(\kappa_{\text{A}}\leftarrow \sigma_{\max}(\mathbf{A})\) and set \(\widetilde{\mathbf{A}}\leftarrow \mathbf{A}/\kappa_{\text{A}}\)
        \STATE Compute SVD \(\widetilde{\mathbf{A}}=\mathbf{U}_{\text{A}}\boldsymbol{\Sigma}\mathbf{V}_{\text{A}}^\mathrm{H}\), extract \(\bar{\boldsymbol{\Sigma}}\), and partition \(\mathbf{V}_{\text{A}}=[\mathbf{V}_{\text{A},1}\ \mathbf{V}_{\text{A},2}]\)
        \STATE \textbf{// Step 2: Left Factor Unitary Synthesis}
        \STATE Synthesize \(\mathbf{U}_{\text{A}}\) via Algorithm~\ref{alg:clements_revised} to obtain \((\mathbf{L}_{\mathrm U}, \mathbf{R}_{\mathrm U}', \mathbf{D}_{\mathrm U}')\)
        \STATE \textbf{// Step 3: Right Factor Semi-Unitary Synthesis}
        \STATE Initialize \(\mathbf{W}\leftarrow \mathbf{V}_{\text{A},1}\)
        \FOR{\(c=1\) to \(M\)}
        \FOR{\(r=N\) down to \(c+1\)}
        \STATE Compute \((\theta_{r,c}^{\mathrm V},\phi_{r,c}^{\mathrm V})\) via Lemma~\ref{thm:row_annihilation}
        \STATE Update \(\mathbf{W}\leftarrow (\mathbf{G}_{r-1}(\theta_{r,c}^{\mathrm V},\phi_{r,c}^{\mathrm V}))^\mathrm{H}\mathbf{W}\)
        \ENDFOR
        \ENDFOR
        \STATE Set \(\mathbf{D}_{\mathrm V} \leftarrow \operatorname{diag} \big( [\mathbf{W}]_{1,1},\ldots,[\mathbf{W}]_{M,M} \big)\) and form \(\mathbf{L}_{\mathrm V}\) by instantiating \eqref{eq:LQ_full} with \(\mathbf{Q}=\mathbf{V}_{\text{A},1}\)
        \STATE \textbf{// Step 4: Standardization and Scaling}
        \STATE Form \(\mathbf{D}_{\mathrm V,N}\) via \eqref{eq:Ds_extension_revised} and initialize \(\mathbf{D}_{\mathrm V,N}'\leftarrow \mathbf{D}_{\mathrm V,N}^\mathrm{H}\)
        \FOR{\(c=M\) down to \(1\)}
        \FOR{\(r=c+1\) to \(N\)}
        \STATE Compute \(\phi_{r,c}^{\prime\mathrm V} \) according to \eqref{eq:embedded_push_phi}
        \STATE Update \(\mathbf{D}_{\mathrm V,N}'\) according to \eqref{eq:embedded_push_d1}--\eqref{eq:embedded_push_d2} with \(p=r-1\)
        \ENDFOR
        \ENDFOR
        \STATE Form the standardized \(\mathbf{R}_{\mathrm V}'\) according to \eqref{eq:Rs_prime_full_revised}
        \STATE Compute \((\theta_m^{\Gamma}, \phi_m^{\Gamma})\) for \(\boldsymbol{\Gamma} = \mathbf{D}_{\mathrm U}'\bar{\boldsymbol{\Sigma}}\) via \eqref{eq:gamma_mzi_theta_revised} and \eqref{eq:gamma_mzi_phi_revised}
        \STATE \textbf{Output:} \(\kappa_{\text{A}}\), all MZI phase configurations for \(\mathbf{L}_{\mathrm U}, \mathbf{R}_{\mathrm U}', \mathbf{R}_{\mathrm V}'\), \(\boldsymbol{\Gamma}\), and \(\mathbf{D}_{\mathrm V,N}'\)
    \end{algorithmic}
\end{algorithm}

Equation~\eqref{eq:proposed_arbitrary_factorization_scaled_revised} corresponds directly to a four-part hardware implementation architecture. The left part, \(\mathbf{L}_{\mathrm U}\mathbf{R}_{\mathrm U}'\), is the standard \(M\times M\) Clements rectangular MZI network produced by the unitary matrix synthesis. The middle part, \(\boldsymbol{\Gamma}\), is an \(M\)-channel singular-value scaling layer with additional phase corrections. The right part, \(\mathbf{R}_{\mathrm V}'\), is realized by standard MZI elements produced by the semi-unitary matrix synthesis. The terminal diagonal layer \(\mathbf{D}_{\mathrm V,N}'\) is implemented by \(N\) single-channel phase shifters at the input boundary. Therefore, compared with the conventional realization that synthesizes both singular matrices as square unitary meshes, the proposed architecture replaces the redundant \(N\times N\) unitary mesh on the larger side by a compact \(N\times M\) semi-unitary realization.

The entries of the middle diagonal scaling layer \(\boldsymbol{\Gamma} = \operatorname{diag}(g_1, \ldots, g_M)\) satisfy \(|g_m| = \sigma_m \le 1\). Each entry can be realized by one standard MZI element operated as a tunable single-channel scaler. Specifically, for the $m$-th MZI element with parameters $(\theta_m^{\Gamma},\phi_m^{\Gamma})$, if the input signal \(x_m\) is injected into the upper input port while the lower input port is left unexcited, the upper output obtained from \eqref{eq:mzi_block_explicit_revised} is
\begin{equation}
    \label{eq:gamma_mzi_output_revised}
    y_m = j e^{j(\theta_m^{\Gamma}/2+\phi_m^{\Gamma})} \sin(\theta_m^{\Gamma}/2)\,x_m.
\end{equation}
Hence, the target coefficient \(g_m\) can be synthesized by choosing
\begin{equation}
    \label{eq:gamma_mzi_theta_revised}
    \theta_m^{\Gamma} = 2\arcsin |g_m|,
\end{equation}
\begin{equation}
    \label{eq:gamma_mzi_phi_revised}
    \phi_m^{\Gamma} =  \operatorname{mod} (\angle g_m - \frac{\theta_m^{\Gamma}}{2} - \frac{\pi}{2}, 2\pi),
\end{equation}
which yields \(y_m=g_m x_m\). The lower output port carries the complementary signal power associated with \(1-|g_m|^{2}\) and is terminated. Consequently, \(\boldsymbol{\Gamma}\) is realized by \(M\) parallel MZI scalers. Note that this single-channel scaler design also directly applies to the implementation of \(\boldsymbol{\Sigma}\) in Case~1 by setting \(\angle g_m = 0\).

Moreover, for the complementary case \(M>N\), there exists
\begin{equation}
    \label{eq:thin_svd_tall_revised}
    \widetilde{\mathbf{A}} = \mathbf{U}_{\text{A},1}\bar{\boldsymbol{\Sigma}}\mathbf{V}_{\text{A}}^\mathrm{H},
\end{equation}
where \(\mathbf{U}_{\text{A},1}\in\mathbb{C}^{M\times N}\) is semi-unitary and \(\mathbf{V}_{\text{A}}\in\mathbb{C}^{N\times N}\) is unitary. We use Theorem~\ref{thm:semi_unitary_synthesis} and Algorithm~\ref{alg:clements_revised} to get
\begin{equation}
    \label{eq:factors_tall_revised}
    \mathbf{U}_{\text{A},1} = \mathbf{L}_{\mathrm U}\mathbf{E}_{M,N}\mathbf{D}_{\mathrm U}, \qquad \mathbf{V}_{\text{A}}^\mathrm{H} = \mathbf{L}_{\mathrm V}\mathbf{R}_{\mathrm V}'\mathbf{D}_{\mathrm V}',
\end{equation}
respectively. Defining \(\boldsymbol{\Gamma} = \mathbf{D}_{\mathrm U}\bar{\boldsymbol{\Sigma}}\), the resulting hardware-consistent factorization is
\begin{equation}
    \label{eq:proposed_arbitrary_factorization_tall_scaled_revised}
    \mathbf{A} = \kappa_{\text{A}} \mathbf{L}_{\mathrm U}\mathbf{E}_{M,N}\boldsymbol{\Gamma}\mathbf{L}_{\mathrm V}\mathbf{R}_{\mathrm V}'\mathbf{D}_{\mathrm V}'.
\end{equation}
Here, all constituent MZI factors are already in the standard form. No additional standardization is required.

\subsection{Hardware and Algorithm Complexity}
Taking into account all constituent stages for synthesizing an arbitrary $M\times N$ matrix $\mathbf{A}$ with $M \le N$, the overall hardware requires $K_{\mathrm U} = M(M-1)/2$ MZI elements for the $M\times M$ unitary mesh $\mathbf{L}_{\mathrm U}\mathbf{R}_{\mathrm U}'$, $K_{\Gamma} = M$ tunable MZI scalers for $\boldsymbol{\Gamma}$, and $K_{\mathrm V} = NM - M(M+1)/2$ MZI elements for the right semi-unitary mesh $\mathbf{R}_{\mathrm V}'$. Summing these contributions yields the total MZI element count:
\begin{equation}
    \label{eq:total_mzi_count}
    K_{\mathrm{prop}} = K_{\mathrm U} + K_{\Gamma} + K_{\mathrm V} = NM.
\end{equation}
Remarkably, \eqref{eq:total_mzi_count} demonstrates that synthesizing an arbitrary $M \times N$ complex operator via the proposed semi-unitary WPU architecture requires exactly $NM$ elementary MZI elements. In comparison, the conventional full-SVD realization requires $K_{\mathrm{full}} = M(M-1)/2 + M + N(N-1)/2 = (M^2+N^2+M-N)/2$ MZI elements. Therefore, the proposed semi-unitary realization eliminates
\begin{equation}
    \label{eq:count_saving_revised}
    K_{\mathrm{full}} - K_{\mathrm{prop}} = \frac{(N-M)(N-M-1)}{2}
\end{equation}
redundant MZI elements. This represents an asymptotic hardware reduction from $\mathcal{O}(N^2+M^2)$ to $\mathcal{O}(NM)$. Furthermore, since each MZI element incorporates two phase shifters ($\theta$ and $\phi$), the $NM$ MZI elements along with the $N$ boundary phase shifters in $\mathbf{D}_{\mathrm V,N}'$ require a total of $2NM + N$ tunable phase controls, which closely matches the $2NM$ real parameters inherent to an arbitrary unconstrained $M \times N$ complex matrix.

The compilation complexity is also reduced. In the semi-unitary elimination step, the update associated with column \(c\) only acts on the remaining nonzero submatrix formed by columns \(c,\ldots,M\). Thus, the arithmetic complexity of the semi-unitary compilation is
\begin{equation}
    \label{eq:semi_compile_revised}
    C_{\mathrm V} = \sum_{c=1}^{M}(N-c)(M-c+1) = \frac{M(M+1)(3N-M-2)}{6},
\end{equation}
which scales as \(\mathcal{O}(NM^{2})\) in terms of complex multiply-accumulate operations. By contrast, compiling the redundant full \(N\times N\) unitary factor scales cubically with \(N\), i.e., \(\mathcal{O}(N^{3})\). Hence, in strongly rectangular settings with \(M\ll N\), the proposed method significantly reduces both the hardware complexity and the closed-form compilation time.

\begin{figure*}[!t]
    \centering
    \includegraphics[width=0.85\textwidth]{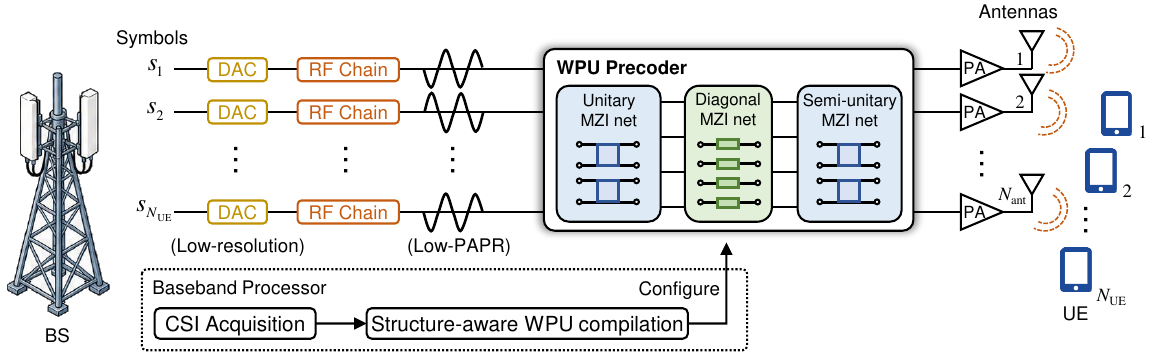}
    \caption{A WPU-assisted multiuser downlink precoding system. The baseband processor performs CSI acquisition and WPU compilation once per coherence block. At symbol transmission time, the WPU transforms the \(N_{\mathrm{UE}}\) data streams into the \(N_{\mathrm{ant}}\) spatial signals in the wave domain.}
    \label{fig:wpu_precoding_architecture}
\end{figure*}

\section{Wave-Domain MultiUser Precoding Design}
\label{sec:precoding}

Matrix–vector multiplication underlies many wireless communication tasks, including beamforming, parameter estimation, and signal detection. The preceding sections established that the WPU can synthesize an arbitrary complex-valued matrix transformation in the wave domain. In this section, we examine precoding as a representative physical-layer application and assess the link-level benefits of implementing this transformation directly in the wave domain.

\subsection{Signal Model}

As illustrated in Fig.~\ref{fig:wpu_precoding_architecture}, we consider a multiuser downlink precoding system, where a base station equipped with \(N_{\mathrm{ant}}\) transmit antennas serves \(N_{\mathrm{UE}}\) single-antenna UEs. The transmitted data streams are first generated in the baseband processor, converted by low-resolution digital-to-analog converters (DACs) and RF chains, and then processed by the WPU precoder into the spatial-domain signals. At the beginning of each channel coherence block, the baseband processor acquires the channel state information (CSI), designs the MMSE precoding matrix, and compiles it into the tunable parameters of the WPU. During the subsequent symbol transmissions within the same coherence block, the configured WPU is reused as a fixed wave-domain precoder.

Denote the downlink channel vector from the base station to UE \(k\) by \(\mathbf{h}_k\in\mathbb{C}^{N_{\mathrm{ant}}\times 1}\), and collect all user channels into \(\mathbf{H}_{\mathrm{dl}}=[\mathbf{h}_1,\ldots,\mathbf{h}_{N_{\mathrm{UE}}}]^{\mathrm H}\in\mathbb{C}^{N_{\mathrm{UE}}\times N_{\mathrm{ant}}}\), with \(N_{\mathrm{ant}}\ge N_{\mathrm{UE}}\). Let \(\mathbf{s}=[s_1,\ldots,s_{N_{\mathrm{UE}}}]^{\mathrm T}\) denote the transmitted data-symbol vector satisfying \(\mathbb{E}[\mathbf{s}\mathbf{s}^{\mathrm H}]=\mathbf{I}_{N_{\mathrm{UE}}}\). Let \(\mathbf{F}=[\mathbf{f}_1,\ldots,\mathbf{f}_{N_{\mathrm{UE}}}]\in\mathbb{C}^{N_{\mathrm{ant}}\times N_{\mathrm{UE}}}\) be the digital precoding matrix, where \(\mathbf{f}_j\) is the spatial precoding vector for UE \(j\).

Under ideal conditions, the WPU can synthesize the exact target precoding matrix \(\mathbf{F}\) in closed form. However, in practical implementations, hardware impairments inevitably introduce insertion loss and phase errors. We model the realized WPU precoder as
\begin{equation}
    \label{eq:wpu_realized_precoder}
    \widehat{\mathbf{F}} = \alpha \left( \mathbf{F} + \Delta\mathbf{F} \right),
\end{equation}
where \(\alpha\in(0,1]\) is the overall attenuation factor of the WPU, and \(\Delta\mathbf{F}=[\mathbf{e}_1,\ldots,\mathbf{e}_{N_{\mathrm{UE}}}]\in\mathbb{C}^{N_{\mathrm{ant}}\times N_{\mathrm{UE}}}\) denotes the residual operator error induced by hardware perturbations such as phase shifter quantization. The transmitted signal across the antenna array is \(\mathbf{x} = \rho \widehat{\mathbf{F}}\mathbf{s}\), where the scalar $\rho = \sqrt{P_{\mathrm T}/\operatorname{tr}(\widehat{\mathbf{F}}\widehat{\mathbf{F}}^{\mathrm H})}$ normalizes the total radiated power to the base station transmit power budget \(P_{\mathrm T}\).

The received signal at UE \(k\) is expressed as
\begin{align}
    y_k &= \underbrace{ \rho\alpha \mathbf{h}_k^{\mathrm H} ( \mathbf{f}_k + \mathbf{e}_k) s_k }_{\text{desired signal}} \notag\\
    &\quad+ \underbrace{ \rho\alpha \mathbf{h}_k^{\mathrm H} \sum_{j\ne k} (\mathbf{f}_j+\mathbf{e}_j) s_j }_{\text{multiuser and hardware-induced interference}} + \underbrace{ n_k }_{\text{thermal noise}},
    \label{eq:wpu_received_signal_expanded}
\end{align}
where \(n_k\sim\mathcal{CN}(0,\sigma_n^2)\) is the additive thermal noise. The desired term includes the implementation error because the receiver is assumed to estimate the effective channel \(\rho\alpha\mathbf{h}_k^{\mathrm H}(\mathbf{f}_k+\mathbf{e}_k)\) rather than only the ideal \(\rho\alpha\mathbf{h}_k^{\mathrm H}\mathbf{f}_k\). The SINR of UE \(k\) is therefore given by
\begin{equation}
    \label{eq:wpu_mu_miso_sinr}
    \gamma_k^{\mathrm{WPU}} = \frac{\rho^2\alpha^2 \left| \mathbf{h}_k^{\mathrm H}\mathbf{f}_k + \mathbf{h}_k^{\mathrm H}\mathbf{e}_k \right|^2 }{\rho^2\alpha^2 \sum_{j\ne k} \left| \mathbf{h}_k^{\mathrm H}\mathbf{f}_j + \mathbf{h}_k^{\mathrm H}\mathbf{e}_j \right|^2 + \sigma_n^2} .
\end{equation}
The achievable sum rate realized by the WPU is
\begin{equation}
    \label{eq:wpu_mu_miso_sum_rate}
    R_{\mathrm{sum}}^{\mathrm{WPU}} = \sum_{k=1}^{N_{\mathrm{UE}}} \log_2 \left( 1+ \gamma_k^{\mathrm{WPU}} \right).
\end{equation}
When \(\Delta\mathbf{F}=\mathbf{0}\), \eqref{eq:wpu_mu_miso_sinr} and \eqref{eq:wpu_mu_miso_sum_rate} reduce to the sum rate of the ideal linear precoder.

\subsection{WPU Precoder Design}

To effectively balance multiuser interference mitigation and noise enhancement, we adopt the MMSE precoding matrix as the target digital precoder, which is given by
\begin{equation}
    \label{eq:mu_miso_mmse_precoder}
    \mathbf{F} = \mathbf{H}_{\mathrm{dl}}^{\mathrm H} \left( \mathbf{H}_{\mathrm{dl}} \mathbf{H}_{\mathrm{dl}}^{\mathrm H} + \mu\mathbf{I}_{N_{\mathrm{UE}}} \right)^{-1},
\end{equation}
with \(\mu>0\) denoting the regularization coefficient. For compiling this matrix onto WPU, a straightforward approach is to explicitly calculate the dense matrix \(\mathbf{F}\in\mathbb{C}^{N_{\mathrm{ant}}\times N_{\mathrm{UE}}}\) in the baseband and then synthesize it using the non-unitary algorithm. However, this method requires matrix inversion in MMSE calculation and a subsequent SVD of \(\mathbf{F}\). To reduce this computational overhead, we exploit the algebraic structure of \(\mathbf{F}\) and the SVD of \(\mathbf{H}_{\mathrm{dl}}\) to establish a more efficient closed-form compilation procedure.

Consider the compact SVD of the downlink channel matrix:
\begin{equation}
    \label{eq:wpu_channel_compact_svd}
    \mathbf{H}_{\mathrm{dl}} = \mathbf{U}_{\text{H}} \boldsymbol{\Lambda}_{\text{H}} \mathbf{V}_{\text{H}}^{\mathrm H},
\end{equation}
where \(\boldsymbol{\Lambda}_{\text{H}} = \operatorname{diag} ( \lambda_1,\ldots,\lambda_{N_{\mathrm{UE}}} )\), \(\mathbf{U}_{\text{H}}\in \mathbb{C}^{N_{\mathrm{UE}}\times N_{\mathrm{UE}}}\) is unitary, and \(\mathbf{V}_{\text{H}}\in \mathbb{C}^{N_{\mathrm{ant}}\times N_{\mathrm{UE}}}\) is semi-unitary, satisfying \(\mathbf{V}_{\text{H}}^{\mathrm H}\mathbf{V}_{\text{H}} = \mathbf{I}_{N_{\mathrm{UE}}}\). Substituting \eqref{eq:wpu_channel_compact_svd} into \eqref{eq:mu_miso_mmse_precoder} yields
\begin{gather}
    \label{eq:wpu_mmse_structured_factorization}
    \mathbf{F} = \mathbf{V}_{\text{H}} \mathbf{G}_{\mu} \mathbf{U}_{\text{H}}^{\mathrm H}, \\ \mathbf{G}_{\mu} = \operatorname{diag} \left( \frac{\lambda_1}{\lambda_1^2+\mu}, \ldots, \frac{\lambda_{N_{\mathrm{UE}}}} {\lambda_{N_{\mathrm{UE}}}^2+\mu} \right).
    \label{eq:wpu_mmse_structured_factorization_Gmu}
\end{gather}
Equation \eqref{eq:wpu_mmse_structured_factorization} indicates that the SVD components of the channel matrix \(\mathbf{H}_{\mathrm{dl}}\) directly provide the constituent factor matrices for expressing $\mathbf{F}$, where the semi-unitary matrix \(\mathbf{V}_{\mathrm H}\) and the unitary matrix \(\mathbf{U}_{\mathrm H}^{\mathrm H}\) can be synthesized individually by the WPU while \(\mathbf{G}_{\mu}\) serves as the diagonal scaling layer. Consequently, WPU synthesis for the MMSE precoder can bypass the matrix inversion and SVD of \(\mathbf{F}\), thereby reducing the synthesis overhead.

\begin{algorithm}[t]
    \caption{Structure-Aware WPU Synthesis for MMSE Precoder}
    \label{alg:wpu_mmse_direct_compilation}
    \begin{algorithmic}[1]
        \STATE \textbf{Input:} Downlink channel \(\mathbf{H}_{\mathrm{dl}} \in \mathbb{C}^{N_{\mathrm{UE}}\times N_{\mathrm{ant}}}\) and regularization coefficient \(\mu\)
        
        \STATE Compute the compact SVD of \(\mathbf{H}_{\mathrm{dl}}\) according to \eqref{eq:wpu_channel_compact_svd} and get ($\mathbf{U}_{\text{H}}, \boldsymbol{\Lambda}_{\text{H}}, \mathbf{V}_{\text{H}}$)
        
        \STATE Compute \({\mathbf{G}}_{\mu}\) according to \eqref{eq:wpu_mmse_structured_factorization_Gmu}
        
        \STATE Compute the normalization factor \(\beta_{\mu}\) according to \eqref{eq:wpu_mmse_passive_normalization} and normalize \({\mathbf{G}}_{\mu}\) into \(\overline{\mathbf{G}}_{\mu}\)
        
        \STATE Synthesize the unitary factor \(\mathbf{U}_{\text{H}}^{\mathrm H}\) via Algorithm~\ref{alg:clements_revised} to obtain \((\mathbf{L}_{\mathrm U}, \mathbf{R}_{\mathrm U}^{\prime}, \mathbf{D}_{\mathrm U}^{\prime})\) and the associated MZI parameters
        
        \STATE Synthesize the semi-unitary factor \(\mathbf{V}_{\mathrm H}\) via Step~3 of Algorithm~\ref{alg:proposed_arbitrary_revised} to obtain \(\mathbf{L}_{\mathrm V}\), \(\mathbf{D}_{\mathrm V}\), and the associated MZI parameters
        
        \STATE Form \(\boldsymbol{\Gamma}_{\mu} \leftarrow \mathbf{D}_{\mathrm V}\overline{\mathbf{G}}_{\mu}\), and map each \([\boldsymbol{\Gamma}_{\mu}]_{m,m}\) to MZI scaler parameters \((\theta_m^{\Gamma}, \phi_m^{\Gamma})\) via \eqref{eq:gamma_mzi_theta_revised} and \eqref{eq:gamma_mzi_phi_revised}
        
        \STATE \textbf{Output:} Global scaling factor \(\beta_{\mu}\) and phase configurations for \(\mathbf{L}_{\mathrm U}, \mathbf{R}_{\mathrm U}^{\prime}, \mathbf{D}_{\mathrm U}^{\prime}, \boldsymbol{\Gamma}_{\mu}\), and \(\mathbf{L}_{\mathrm V}\)
    \end{algorithmic}
\end{algorithm}

Algorithm~\ref{alg:wpu_mmse_direct_compilation} summarizes the WPU synthesis procedure of MMSE precoder. To satisfy the passivity requirement of the WPU, we normalize \(\mathbf{F}\) by defining
\begin{equation}
    \label{eq:wpu_mmse_passive_normalization}
    \beta_{\mu} = \max_{i\in\{1,\ldots,N_{\mathrm{UE}}\}} \frac{\lambda_i}{\lambda_i^2+\mu},
\end{equation}
and $\overline{\mathbf{F}} = {\mathbf{F}}/{\beta_{\mu}} = \mathbf{V}_{\text{H}} \overline{\mathbf{G}}_{\mu} \mathbf{U}_{\text{H}}^{\mathrm H},$ where $\overline{\mathbf{G}}_{\mu} = {\mathbf{G}_{\mu}}/{\beta_{\mu}}$
with \(\bar g_i=[\overline{\mathbf G}_{\mu}]_{i,i}\in[0,1]\). Because the precoding matrix corresponds to the \(M > N\) tall-matrix case in Section~\ref{sec:non_unitary_synthesis}, $\overline{\mathbf{F}}$ is compiled into the WPU via the semi-unitary synthesis of \(\mathbf{V}_{\text{H}}\) and the unitary synthesis of \(\mathbf{U}_{\text{H}}^{\mathrm H}\). Applying Theorem~\ref{thm:semi_unitary_synthesis} directly to the semi-unitary matrix \(\mathbf{V}_{\text{H}}\in\mathbb{C}^{N_{\mathrm{ant}}\times N_{\mathrm{UE}}}\) produces \(\mathbf{V}_{\text{H}} = \mathbf{L}_{\mathrm V} \mathbf{E}_{N_{\mathrm{ant}},N_{\mathrm{UE}}} \mathbf{D}_{\mathrm V}\), where \(\mathbf{L}_{\mathrm V}\) is an \(N_{\mathrm{ant}}\times N_{\mathrm{ant}}\) mesh formed by \(K_{\mathrm V} = N_{\mathrm{ant}}N_{\mathrm{UE}} - N_{\mathrm{UE}}(N_{\mathrm{UE}}+1)/2\) standard MZI elements, and \(\mathbf{D}_{\mathrm V} = \operatorname{diag}(d_1,\ldots,d_{N_{\mathrm{UE}}})\) is a diagonal phase matrix. Applying Algorithm~\ref{alg:clements_revised} to \(\mathbf{U}_{\text{H}}^{\mathrm H}\in\mathbb{C}^{N_{\mathrm{UE}}\times N_{\mathrm{UE}}}\) yields \(\mathbf{U}_{\text{H}}^{\mathrm H} = \mathbf{L}_{\mathrm U} \mathbf{R}_{\mathrm U}^{\prime} \mathbf{D}_{\mathrm U}^{\prime}\). Defining \(\boldsymbol{\Gamma}_{\mu} = \mathbf{D}_{\mathrm V} \overline{\mathbf{G}}_{\mu}\), the passive hardware realization of \(\overline{\mathbf{F}}\) is factorized as
\begin{equation}
    \label{eq:wpu_mmse_hardware_consistent_factorization}
    \overline{\mathbf{F}} = \mathbf{L}_{\mathrm V} \mathbf{E}_{N_{\mathrm{ant}},N_{\mathrm{UE}}} \boldsymbol{\Gamma}_{\mu} \mathbf{L}_{\mathrm U} \mathbf{R}_{\mathrm U}^{\prime} \mathbf{D}_{\mathrm U}^{\prime}.
\end{equation}

Algorithm~\ref{alg:wpu_mmse_direct_compilation} specializes the general matrix-synthesis framework to MMSE precoding. Instead of constructing \(\mathbf{F}\) and treating it as an unstructured rectangular matrix, the algorithm reuses the singular vectors of the downlink channel and updates the singular mode gains according to \eqref{eq:wpu_mmse_structured_factorization_Gmu}. Once configured, the resulting WPU implements the normalized MMSE transformation during each symbol interval.

\begin{table*}[!t]
    \caption{Comparison of Precoding Architectures in MultiUser MISO Systems}
    \label{tab:precoding_comparison}
    \centering
    \renewcommand{\arraystretch}{1.25}
    \begin{tabular}{lccccc}
        \toprule
        \textbf{Architecture} & \textbf{Spatial DoF} & \textbf{RF Chains ($N_{\mathrm{RF}}$)} & \textbf{MVM Complexity} & \textbf{Configuration Method} & \textbf{Configuration Complexity} \\
        \midrule
        \textbf{Fully Digital} & Full & $N_{\mathrm{ant}}$ & $\mathcal{O}(N_{\mathrm{ant}}N_{\mathrm{UE}})$ & Closed-form & $\mathcal{O}(N_{\mathrm{UE}}^3 + N_{\mathrm{ant}}N_{\mathrm{UE}}^2)$ \\
        \textbf{Analog (Codebook)} & Highly constrained & $N_{\mathrm{UE}}$ & $\mathcal{O}(1)$ & Codebook search & $\mathcal{O}(N_{\mathrm{UE}} N_{\mathrm{code}})$ \\
        \textbf{Hybrid (PZF)} \cite{liang2014hybrid} & Partially constrained & $N_{\mathrm{UE}}\le N_{\mathrm{RF}} < N_{\mathrm{ant}}$ & $\mathcal{O}(N_{\mathrm{RF}}N_{\mathrm{UE}})$ & Closed-form & $\mathcal{O}(N_{\mathrm{ant}}N_{\mathrm{UE}}^2 + N_{\mathrm{UE}}^3)$ \\
        \textbf{SIM} \cite{an2025twc} & Partially constrained & $N_{\mathrm{UE}}$ & $\mathcal{O}(1)$ & Iterative optimization & $\mathcal{O}(I_{\mathrm{iter}} L N_{\mathrm{meta}}^2 N_{\mathrm{UE}})$ \\
        \textbf{WPU (Proposed)} & \textbf{Full} & \textbf{$\bm{N_{\mathrm{UE}}}$} & $\bm{\mathcal{O}(1)}$ & \textbf{Closed-form} & $\bm{\mathcal{O}(N_{\mathrm{ant}}N_{\mathrm{UE}}^2)}$ \\
        \bottomrule
    \end{tabular}
\end{table*}

\subsection{Architectural Comparison and Analysis}
\label{subsec:wpu_latency_energy}

To contextualize the advantages of the proposed WPU, Table~\ref{tab:precoding_comparison} compares its architecture with fully digital, analog, hybrid, and SIM precoding schemes in terms of spatial DoF, hardware overhead, and computational characteristics. As summarized in Table~\ref{tab:precoding_comparison}, the proposed WPU exhibits architectural superiority across the evaluated metrics. First, while analog and hybrid architectures sacrifice spatial DoF due to discrete codebooks or constant-modulus phase constraints, and SIM is restricted by fixed inter-layer diffraction geometry, the WPU achieves full spatial DoF with the minimal RF chain count $N_{\mathrm{RF}} = N_{\mathrm{UE}}$ by programming cascaded planar MZI elements. Second, unlike digital and hybrid schemes that incur $\mathcal{O}(N_{\mathrm{ant}}N_{\mathrm{UE}})$ or $\mathcal{O}(N_{\mathrm{RF}}N_{\mathrm{UE}})$ baseband multiply-accumulate operations per symbol vector, the WPU executes the spatial transformation through light-speed EM wave propagation, achieving $\mathcal{O}(1)$ digital arithmetic complexity with extremely low latency. Third, while SIM relies on non-convex iterative optimization (e.g., alternating optimization or projected gradient descent) that incurs high configuration latency and risks local convergence, the WPU synthesis determines all hardware parameters via a closed-form algorithm scaling as $\mathcal{O}(N_{\mathrm{ant}}N_{\mathrm{UE}}^2)$. Consequently, the WPU provides a balanced configuration-and-computing latency tradeoff for real-time spatial processing.

\section{Numerical Results}
\label{sec:numerical_results}

In this section, we present numerical results to evaluate the proposed WPU architecture, closed-form matrix synthesis, and precoding framework across both operator-level and link-level metrics.

\subsection{Closed-Form Matrix Synthesis Performance}
\label{subsec:matrix_synthesis_perf}

This subsection evaluates the matrix synthesis accuracy and overhead of the proposed WPU framework under an ideal model utilized in Sections~\ref{sec:architecture} and \ref{sec:non_unitary_synthesis}. The simulations are performed on an Intel Core i7-12700 CPU at $2.10$~GHz using double-precision complex arithmetic. The normalized mean-square error (NMSE) between the target operator $\mathbf{A}$ and its synthesized counterpart $\widehat{\mathbf{A}}$ is adopted as the primary metric:
\begin{equation}
    \mathrm{NMSE} = \frac{\|\widehat{\mathbf{A}}-\mathbf{A}\|_{\mathrm{F}}^{2}}{\|\mathbf{A}\|_{\mathrm{F}}^{2}}.
\end{equation}
For general complex-valued operators, the entries of $\mathbf{A}$ are generated as independent and identically distributed (i.i.d.) circularly symmetric complex Gaussian variables. 

\begin{table*}[!t]
    \caption{Synthesis Accuracy, Runtime, and Hardware Complexity for General Complex-Valued Square Operators}
    \label{tab:wave_arch_synthesis}
    \centering
    \renewcommand{\arraystretch}{1.2}
    \setlength{\tabcolsep}{14pt}
    \footnotesize
    \begin{tabular}{llccc}
        \toprule
        Dimension & Architecture & Mean NMSE & Runtime (s) & Tunable Parameters \\
        \midrule
        \multirow{5}{*}{$4\times 4$} 
        & \textbf{WPU (Proposed)}           & $\mathbf{2.77\times 10^{-30}}$ & $\mathbf{0.001}$ & $\mathbf{36}$ \\
        & SIM (Balanced)                   & $1.79\times 10^{-3}$          & $0.655$          & $36$   \\
        & SIM ($\approx 4\times$ Params)   & $4.68\times 10^{-11}$         & $0.762$          & $144$  \\
        & PDNN (Balanced)                  & $9.07\times 10^{-3}$          & $0.534$          & $36$   \\
        & PDNN ($\approx 4\times$ Params)  & $6.80\times 10^{-11}$         & $0.785$          & $144$  \\
        \midrule
        \multirow{5}{*}{$16\times 16$} 
        & \textbf{WPU (Proposed)}           & $\mathbf{3.43\times 10^{-29}}$ & $\mathbf{0.006}$ & $\mathbf{528}$ \\
        & SIM (Balanced)                   & $2.17\times 10^{-3}$          & $1.088$          & $528$  \\
        & SIM ($\approx 4\times$ Params)   & $4.03\times 10^{-11}$         & $3.585$          & $2304$ \\
        & PDNN (Balanced)                  & $6.80\times 10^{-2}$          & $1.140$          & $528$  \\
        & PDNN ($\approx 4\times$ Params)  & $1.15\times 10^{-10}$         & $3.275$          & $2112$ \\
        \midrule
        \multirow{5}{*}{$64\times 64$} 
        & \textbf{WPU (Proposed)}           & $\mathbf{4.46\times 10^{-28}}$ & $\mathbf{0.116}$ & $\mathbf{8256}$ \\
        & SIM (Balanced)                   & $6.43\times 10^{-3}$          & $44.786$         & $8256$ \\
        & SIM ($\approx 4\times$ Params)   & $7.47\times 10^{-10}$         & $356.291$        & $33856$\\
        & PDNN (Balanced)                  & $2.26\times 10^{-3}$          & $44.857$         & $8256$ \\
        & PDNN ($\approx 4\times$ Params)  & $7.38\times 10^{-11}$         & $331.843$        & $33024$\\
        \bottomrule
    \end{tabular}
\end{table*}


Table~\ref{tab:wave_arch_synthesis} summarizes the synthesis accuracy, runtime, and parameter requirements for $4\times4$, $16\times16$, and $64\times64$ operators, averaged over five independent trials. The baselines are stacked intelligent metasurfaces (SIM) based on Rayleigh--Sommerfeld diffraction \cite{an2023jsac} and planar diffractive neural networks (PDNN) based on branchline-coupler and phase-shifter cascades \cite{teng2025pdnn}. To establish a fair comparison, the balanced configurations strictly align their total number of tunable phase shifters with the WPU control budget ($2N^2+N$ phase controls for an $N\times N$ operator). In addition, $4\times$-expanded configurations are evaluated to assess the performance of diffractive networks under substantially augmented hardware resources. All diffractive baselines are optimized via projected gradient descent (PGD) with Adam, an initial learning rate of $0.1$, and 200 iterations.

As reported in Table~\ref{tab:wave_arch_synthesis}, the proposed WPU achieves deterministic matrix synthesis across all evaluated dimensions, with mean NMSEs ranging from $2.77\times10^{-30}$ to $4.46\times10^{-28}$, consistent with numerical errors in double-precision arithmetic. Crucially, the closed-form compilation of the WPU requires only $1$~ms for $4\times4$, $6$~ms for $16\times16$, and $0.116$~s for the large-scale $64\times64$ operator. In contrast, iterative diffractive baselines incur severe performance and computational penalties. Under identical phase-shifter budgets, balanced SIM and PDNN suffer from residual NMSEs between $1.79\times10^{-3}$ and $6.80\times10^{-2}$, failing to reconstruct general complex operators due to the structural rigidity of fixed diffraction or directional-coupling geometries. Quadrupling their hardware budget reduces the NMSE to the order of $10^{-11}$ to $10^{-10}$. However, this gain comes at the cost of prohibitive computational overhead. For the $64\times64$ case, configuring a single matrix requires nearly $45$~s for balanced baselines and exceeds $330$~s for expanded baselines. This represents an enormous compilation speedup of approximately $380\times$ and $3000\times$ for the proposed WPU over balanced and expanded iterative baselines, respectively, demonstrating that the deterministic MZI synthesis resolves the configuration latency bottleneck that plagues optimization-based wave computing.

\begin{figure}[!t]
    \centering
    \includegraphics[height=0.717\columnwidth]{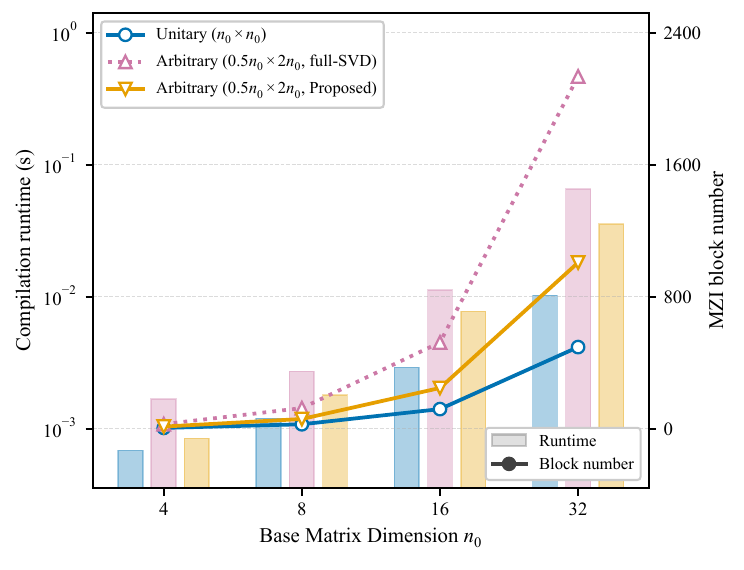}
    \vspace{-0.2cm}
    \caption{Compilation runtime and tunable MZI element count versus base matrix dimension $n_0$.}
    \label{fig:block_runtime_scaling}
    \vspace{-0.3cm}
\end{figure}

Fig.~\ref{fig:block_runtime_scaling} evaluates the hardware overhead and compilation efficiency of the proposed semi-unitary synthesis against the trivial full-SVD approach  as $n_0$ scales from $4$ to $32$. For rectangular operators with dimension $(0.5n_0) \times (2n_0)$, although both closed-form schemes attain exact matrix reconstruction, the proposed synthesis framework avoids completing the right singular factor into a redundant $(2n_0) \times (2n_0)$ unitary matrix. Consequently, the total number of required MZI elements is substantially reduced, scaling as $NM$ in accordance with \eqref{eq:total_mzi_count} and eliminating $(N-M)(N-M-1)/2$ redundant elements as predicted by \eqref{eq:count_saving_revised}. For a $16\times64$ operator, the total MZI count, including the scaling stage, decreases from $2152$ to $1024$, a $52.4\%$ reduction relative to full-SVD synthesis. The hardware reduction also translates into a noticeable decrease in closed-form compilation time. These results confirm the hardware and algorithm efficiency of the proposed architecture for large-scale rectangular operators.

\subsection{Downlink Multiuser Precoding Performance}
\label{subsec:precoding_performance}

This subsection investigates the link-level achievable sum-rate performance of the proposed WPU-assisted precoding scheme in a downlink multiuser MISO system operating over spatially correlated channels. The base station serves $N_{\mathrm{UE}}=8$ single-antenna users under the assumption of perfect CSI and ideal wave-processing hardware, characterized by unity power attenuation ($\alpha=1$) and zero operator implementation perturbation ($\Delta\mathbf{F}=\mathbf{0}$) in \eqref{eq:wpu_realized_precoder}. The spatially correlated channel is synthesized via the transmit-side Kronecker model, given by $\mathbf{H}_{\mathrm{dl}} = \mathbf{G} \mathbf{R}_{\mathrm{t}}^{1/2}$, where the entries of $\mathbf{G} \in \mathbb{C}^{N_{\mathrm{UE}} \times N_{\mathrm{ant}}}$ are independent and identically distributed as $\mathcal{CN}(0,1)$. The spatial correlation matrix is parameterized as $[\mathbf{R}_{\mathrm{t}}]_{m,n} = \rho_{\mathrm{c}}^{|m-n|}$, where $\rho_{\mathrm{c}} \in [0, 1]$ denotes the spatial correlation coefficient. Each row of $\mathbf{H}_{\mathrm{dl}}$ is normalized to satisfy $\|\mathbf{h}_k\|^2 = N_{\mathrm{ant}}$. The transmit SNR is defined as $P_{\mathrm{T}} / \sigma_n^2$ with total radiated power fixed at $P_{\mathrm{T}} = 1$. All competing schemes share identical channel realizations, noise variance, and total power budgets. Numerical results are obtained by averaging over 20 independent channel realizations.

The following representative precoding architectures are benchmarked:
\begin{enumerate}
    \item \emph{Fully Digital Precoding:} The benchmark executes the conventional linear MMSE precoder in \eqref{eq:mu_miso_mmse_precoder} with the standard regularization parameter $\mu = N_{\mathrm{UE}} \sigma_n^2 / P_{\mathrm{T}}$, implemented in the digital baseband with $N_{\mathrm{RF}} = N_{\mathrm{ant}}$ dedicated RF chains.
    \item \emph{Proposed WPU Precoder:} The proposed scheme processes the channel matrix directly into the WPU phase parameters following Algorithm~\ref{alg:wpu_mmse_direct_compilation}, employing only $N_{\mathrm{RF}} = N_{\mathrm{UE}}$ RF chains while completely executing the matrix transformation in the wave domain.
    \item \emph{Hybrid Beamforming (Hybrid BF):} Equipped with $N_{\mathrm{RF}} = N_{\mathrm{UE}}$ RF chains, this baseline employs an analog phase-shifter network matching the conjugate channel phases, concatenated with low-dimensional zero-forcing baseband processing on the resulting effective channel \cite{liang2014hybrid}. 
    \item \emph{Iterative Wave-Domain Baselines (SIM and PDNN):} SIM \cite{an2023jsac} and PDNN \cite{teng2025pdnn} fit the normalized MMSE precoder $\overline{\mathbf{F}}$ via free-space diffraction and directional coupling models, respectively, with phase-shifter budgets matched to the WPU within 10\%. The network weights are updated via projected gradient descent (PGD) with the Adam optimizer for 800 iterations, initialized with a learning rate of $0.1$ and a decay factor of $0.99$.
\end{enumerate}

\begin{figure}[!t]
    \centering
    \includegraphics[width=\columnwidth]{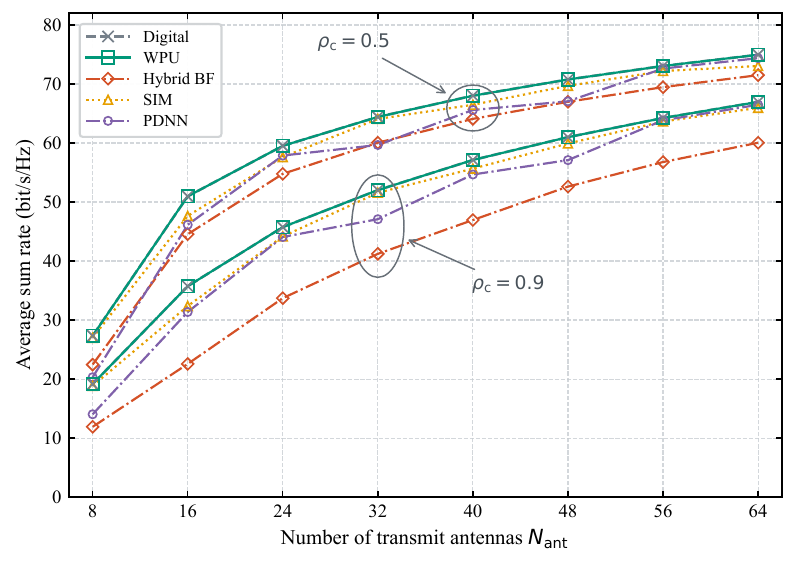}
    \vspace{-0.7cm}
    \caption{Average downlink sum rate versus the number of transmit antennas $N_{\mathrm{ant}}$ for $N_{\mathrm{UE}}=8$ single-antenna users at a transmit SNR of $20$~dB.}
    \vspace{-0.3cm}
    \label{fig:precoding_sumrate_antennas}
\end{figure}

Fig.~\ref{fig:precoding_sumrate_antennas} illustrates the average sum rate versus $N_{\mathrm{ant}}$ at an operating SNR of $20$~dB under moderate ($\rho_{\mathrm{c}} = 0.5$) and severe ($\rho_{\mathrm{c}} = 0.9$) spatial correlation. Across all antenna dimensions and correlation levels, the WPU curves consistently coincide with the digital MMSE benchmark, confirming that the closed-form algorithm realizes wave-domain linear precoding without structural approximation error. In contrast, the remaining architectures exhibit noticeable performance degradations governed by different structural and algorithmic limitations. With $N_{\mathrm{RF}} = N_{\mathrm{UE}}$, hybrid BF falls short of the digital MMSE benchmark, and its low-complexity algorithm incurs more rate losses under strong spatial correlation due to insufficient interference suppression. Although SIM and PDNN outperform Hybrid BF, they still fail to attain the digital benchmark due to residual operator fitting errors, while their iterative optimization incurs compilation overhead two to three orders of magnitude greater than that of WPU.

\begin{figure}[!t]
    \centering
    \includegraphics[width=\columnwidth]{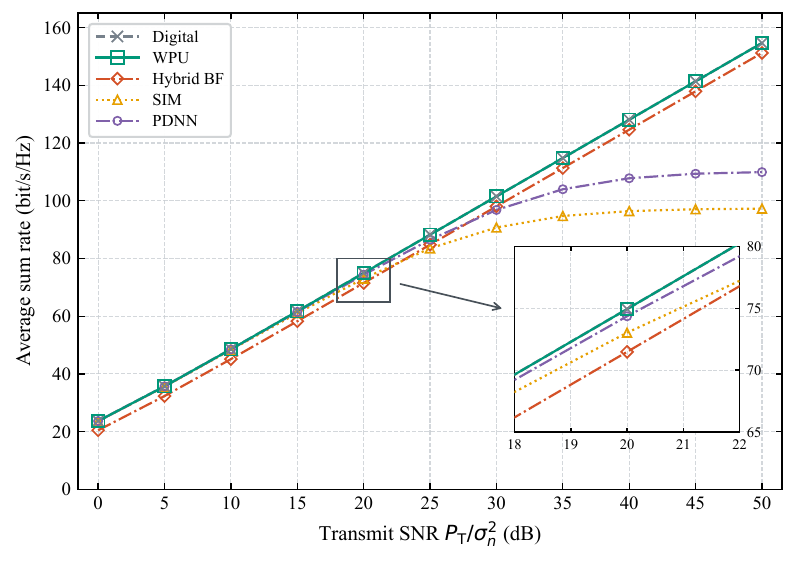}
    \vspace{-0.7cm}
    \caption{Average downlink sum rate versus transmit SNR for $N_{\mathrm{ant}}=64$, $N_{\mathrm{UE}}=8$, and spatial correlation coefficient $\rho_{\mathrm{c}}=0.5$.}
    \vspace{-0.3cm}
    \label{fig:precoding_sumrate_snr}
\end{figure}

Fig.~\ref{fig:precoding_sumrate_snr} evaluates the average sum rate versus the transmit SNR ranging from $0$ to $50$~dB, with $N_{\mathrm{ant}}=64$, $N_{\mathrm{UE}}=8$, and $\rho_{\mathrm{c}}=0.5$. Across the entire SNR sweep, the WPU curves track the digital benchmark with near-zero discrepancy, reaching an average sum rate of $154.65$~bit/s/Hz at $50$~dB. In contrast, SIM and PDNN exhibit interference-limited rate ceilings in the high-SNR regime. This behavior reflects the transition from a noise-limited to an interference-limited regime. As thermal noise becomes negligible, residual operator-fitting errors induce persistent inter-user interference, causing their sum rates to saturate at $97.26$ and $109.92$~bit/s/Hz, respectively. Meanwhile, Hybrid BF retains a non-saturating rate slope via its digital baseband zero-forcing, yet sustains a persistent rate gap of approximately $3.46$~bit/s/Hz relative to the WPU due to the effective-channel gain loss under constant-modulus analog phase constraints.

\subsection{Performance Under Hardware Impairments}
\label{subsec:hardware_robustness}

This subsection evaluates the sum-rate sensitivity to the residual operator error $\Delta\mathbf{F}$ in \eqref{eq:wpu_realized_precoder}. Since the passive attenuation factor $\alpha$ can be compensated by downstream power amplification $\rho$, we set $\alpha=1$ and focus on the impact of operator perturbation. We retain the $64\times8$ precoding setup, spatial correlation coefficient $\rho_{\mathrm{c}}=0.5$, and benchmark configurations of Subsection~\ref{subsec:precoding_performance}. The additive perturbation on the effective precoding matrix $\mathbf{F}$ is modeled as $\Delta\mathbf{F}=\varepsilon\|\mathbf{F}\|_{\mathrm{F}}\mathbf{Z}/\|\mathbf{Z}\|_{\mathrm{F}}$, where $\mathbf{Z}$ has independent and identically distributed entries $Z_{ij}\sim\mathcal{CN}(0,1)$, and $\varepsilon\in\{0.01,0.1\}$ denotes the relative Frobenius error. For baseline architectures, identical relative perturbations are directly applied to their respective realized precoding matrices. The receiver noise variance is normalized to unity, and the transmit power is swept from $0$ to $50$~dB. For each of the 20 channel realizations, 50 independent perturbations are evaluated.

\begin{figure}[!t]
    \centering
    \includegraphics[width=\columnwidth]{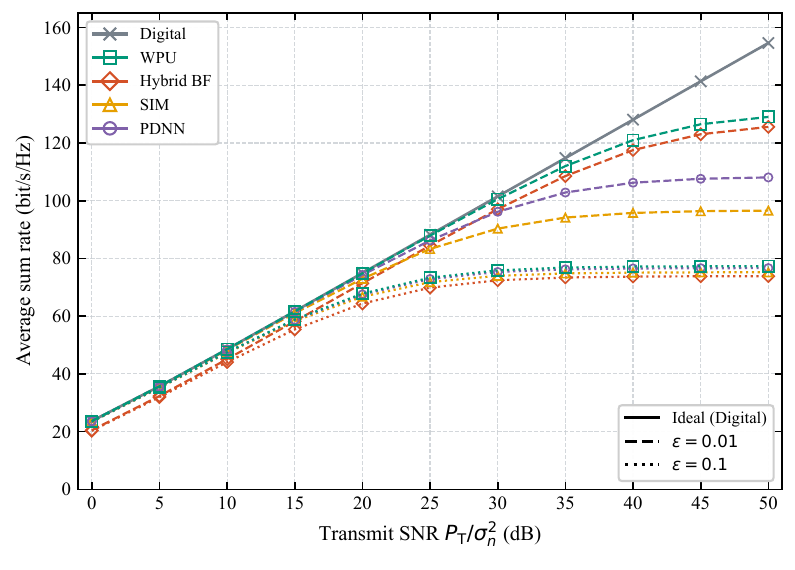}
    \vspace{-0.7cm}
    \caption{Average downlink sum rate under relative Frobenius error $\varepsilon=\|\Delta\mathbf{F}\|_{\mathrm F}/\|\mathbf{F}\|_{\mathrm F}$ for $N_{\mathrm{ant}}=64$, $N_{\mathrm{UE}}=8$, and $\rho_{\mathrm c}=0.5$.}
    \label{fig:matrix_perturbation_sumrate}
    \vspace{-0.3cm}
\end{figure}

Fig.~\ref{fig:matrix_perturbation_sumrate} illustrates the resulting average sum rates for WPU, hybrid BF, SIM, and PDNN across both perturbation levels, with fully digital MMSE precoding serving as the unperturbed benchmark. At $\varepsilon=0.01$, the WPU closely tracks the digital benchmark in the low-to-moderate SNR regime and consistently outperforms the baseline architectures. Therefore, the accuracy advantage of closed-form synthesis remains evident at a relative operator error of $0.01$, although the gap to the fully digital reference grows at high SNR. Increasing the relative error to $\varepsilon=0.1$ severely degrades the high-SNR sum rates and narrows the differences among the architectures. These observations demonstrate that hardware nonidealities degrade achievable sum rate primarily through interference leakage in the interference-limited regime. Consequently, the tolerable hardware imperfections and required control precision are fundamentally governed by the operational SNR range and the acceptable rate loss.

\section{Conclusion}
\label{sec:conclusion}

This paper developed a programmable planar WPU for accelerating wireless physical-layer matrix operators through an MZI-based wave interference network. Starting from the transfer model of a tunable symmetric MZI element, we established a closed-form synthesis procedure for arbitrary unitary transformations and extended it to general complex-valued matrices through SVD and semi-unitary synthesis. The latter eliminates redundant synthesis on the larger matrix dimension, reducing both the hardware and compilation complexity. For multiuser downlink transmission, the general framework was further specialized to MMSE precoding by directly synthesizing the SVD factors of the channel, enabling wave-domain precoding without iterative optimization. Numerical evaluations confirm an NMSE of $4.46\times10^{-28}$ for $64\times64$ matrix synthesis. The proposed framework reduces the total MZI count by $52.4\%$ relative to full-SVD synthesis for $16\times64$ operators and requires $87.5\%$ fewer RF chains than the fully digital architecture for $64$-antenna, $8$-user precoding while matching its sum rate under ideal hardware. Future work will extend the WPU toward programmable wave-domain transceivers for broadband and large-scale MIMO systems, explore its applications to signal detection, channel estimation, and integrated sensing and communication, and validate the WPU design on a reconfigurable RF prototype.

\appendices
\section{Proofs of Lemmas~\ref{thm:row_annihilation} and~\ref{thm:col_annihilation}}
\label{app:row_annihilation}
\label{app:col_annihilation}
Let $s_\theta=\sin(\theta/2)$ and $c_\theta=\cos(\theta/2)$. For row elimination, substituting \eqref{eq:mzi_block_explicit_revised} into $\mathbf{W}'=\mathbf{G}_{r-1}^{\mathrm H}(\theta,\phi)\mathbf{W}$ gives
\begin{equation}
    \label{eq:app_row_target_entry}
    [\mathbf{W}']_{r,c}=-j e^{-j\theta/2}
    \left(e^{-j\phi}c_\theta[\mathbf{W}]_{r-1,c}-s_\theta[\mathbf{W}]_{r,c}\right).
\end{equation}
Write $[\mathbf{W}]_{r-1,c}=a e^{j\psi_1}$ and $[\mathbf{W}]_{r,c}=b e^{j\psi_2}$, with $a,b\ge0$. Choosing $\phi=\psi_1-\psi_2$ and $\tan(\theta/2)=a/b$ makes the parenthesized term $e^{j\psi_2}(ac_\theta-bs_\theta)=0$. For $b=0$, setting $\theta=\pi$ gives the same result.

For column elimination, $\mathbf{W}'=\mathbf{W}\mathbf{G}_{c}(\theta,\phi)$ similarly gives
\begin{equation}
    \label{eq:app_col_target_entry}
    [\mathbf{W}']_{r,c}=j e^{j\theta/2}
    \left(e^{j\phi}s_\theta[\mathbf{W}]_{r,c}+c_\theta[\mathbf{W}]_{r,c+1}\right).
\end{equation}
Write \([\mathbf{W}]_{r,c}=a e^{j\psi_1}\) and \([\mathbf{W}]_{r,c+1}=b e^{j\psi_2}\), where \(a,b\ge0\). The choices $\phi=\psi_2-\psi_1+\pi$ and $\tan(\theta/2)=b/a$ reduce the parenthesized term to $e^{j\psi_2}(-as_\theta+bc_\theta)=0$. For $a=0$, setting $\theta=\pi$ again gives zero. This completes the proofs.

\section{Proof of Lemma~\ref{thm:embedded_push}}
\label{app:diagonal_push}

Partition the original diagonal unitary matrix as
\begin{equation}
    \label{eq:appendix_delta_partition}
    \boldsymbol{\Delta} = \blkdiag\!\big( \boldsymbol{\Delta}_{\mathrm{L}}, \boldsymbol{\Delta}_{\mathrm{loc}}, \boldsymbol{\Delta}_{\mathrm{R}} \big),
\end{equation}
where \(\boldsymbol{\Delta}_{\mathrm{loc}}=\operatorname{diag}(d_p,d_{p+1})\), while \(\boldsymbol{\Delta}_{\mathrm{L}}\) and \(\boldsymbol{\Delta}_{\mathrm{R}}\) contain the unaffected diagonal entries. Since \(\mathbf{G}_{p}\) acts only on coordinates \(p\) and \(p+1\),
\begin{equation}
    \boldsymbol{\Delta} \cdot \left( \mathbf{G}_{p}(\theta,\phi) \right)^\mathrm{H} = \blkdiag\!\big( \boldsymbol{\Delta}_{\mathrm{L}}, \boldsymbol{\Delta}_{\mathrm{loc}} \mathbf{M}^\mathrm{H}(\theta,\phi), \boldsymbol{\Delta}_{\mathrm{R}} \big).
    \label{eq:appendix_embedded_localization}
\end{equation}
Write \(d_p=e^{j\psi_1}\) and \(d_{p+1}=e^{j\psi_2}\). From \eqref{eq:mzi_block_explicit_revised},
\begin{equation}
    \boldsymbol{\Delta}_{\mathrm{loc}} \mathbf{M}^\mathrm{H}(\theta,\phi) = -j e^{-j\theta/2}
    \begin{bmatrix}
        e^{j(\psi_1-\phi)}s_\theta & e^{j\psi_1}c_\theta\\
        e^{j(\psi_2-\phi)}c_\theta & -e^{j\psi_2}s_\theta
    \end{bmatrix},
    \label{eq:appendix_local_left}
\end{equation}
where $s_\theta=\sin(\theta/2)$ and $c_\theta=\cos(\theta/2)$. Let $\phi'=\psi_1-\psi_2$ and construct $\boldsymbol{\Delta}_{\mathrm{loc}}'=\operatorname{diag}(d_p',d_{p+1}')$, where $d_p'=-e^{j(\psi_2-\phi-\theta)}$ and $d_{p+1}'=-e^{j(\psi_2-\theta)}$. Computing \(\mathbf{M}(\theta,\phi')\boldsymbol{\Delta}_{\mathrm{loc}}'\) reproduces \eqref{eq:appendix_local_left}; hence
\begin{equation}
    \label{eq:appendix_local_push_embedded}
    \boldsymbol{\Delta}_{\mathrm{loc}} \mathbf{M}^\mathrm{H}(\theta,\phi) = \mathbf{M}(\theta,\phi') \boldsymbol{\Delta}_{\mathrm{loc}}'.
\end{equation}
Set $\boldsymbol{\Delta}'=\blkdiag(\boldsymbol{\Delta}_{\mathrm{L}},\boldsymbol{\Delta}_{\mathrm{loc}}',\boldsymbol{\Delta}_{\mathrm{R}})$. Substituting \eqref{eq:appendix_local_push_embedded} into \eqref{eq:appendix_embedded_localization} gives
\begin{equation}
    \label{eq:appendix_global_push}
    \boldsymbol{\Delta} \cdot \left( \mathbf{G}_{p}(\theta,\phi) \right)^\mathrm{H} = \mathbf{G}_{p}(\theta,\phi') \boldsymbol{\Delta}'.
\end{equation}
The proof is complete.
\section{Proof of Theorem~\ref{thm:semi_unitary_synthesis}}
\label{app:semi_unitary_synthesis}

Let $\mathbf{Q}\in\mathbb{C}^{N\times M}$ satisfy $\mathbf{Q}^\mathrm{H}\mathbf{Q}=\mathbf{I}_{M}$, and initialize $\mathbf{W}=\mathbf{Q}$. The proof of decomposition starts by applying the adjacent-row elimination procedure in Step~3 of Algorithm~\ref{alg:proposed_arbitrary_revised} to $\mathbf{W}$. This procedure eliminates entries column by column. At each step, Lemma~\ref{thm:row_annihilation} annihilates current $[\mathbf{W}]_{r,c}$ by mixing rows $r-1$ and $r$. Locality preserves the previously established zeros. After processing $c=1,\ldots,M$, the final working matrix therefore has the form
\begin{equation}
    \label{eq:semi_appendix_trapezoid}
    \mathbf{W}_{\mathrm f} =
    \begin{bmatrix}
        \mathbf{T}\\
        \mathbf{0}_{(N-M)\times M}
    \end{bmatrix},
\end{equation}
where $\mathbf{T}\in\mathbb{C}^{M\times M}$ is upper triangular.

Every elimination update is unitary, so the semi-unitary property from $\mathbf{Q}$ is preserved, i.e., $\mathbf{W}_{\mathrm f}^\mathrm{H}\mathbf{W}_{\mathrm f}=\mathbf{I}_{M}$. Substituting \eqref{eq:semi_appendix_trapezoid} into this gives $\mathbf{T}^\mathrm{H}\mathbf{T} =\mathbf{I}_{M}.$ Hence $\mathbf{T}$ is an upper-triangular unitary matrix and must be diagonal. Write $\mathbf{D}_{\mathrm Q}=\mathbf{T}=\operatorname{diag}(d_1,\ldots,d_M)$, where $|d_i|=1$. Equation~\eqref{eq:semi_appendix_trapezoid} then becomes $\mathbf{W}_{\mathrm f}=\mathbf{E}_{N,M}\mathbf{D}_{\mathrm Q}$.

Finally, we reconstruct \(\mathbf{Q}\) from the recorded elimination matrices. Define
\begin{equation}
    \label{eq:semi_appendix_LQ}
    \mathbf{L}_{\mathrm Q} = \prod_{c=1}^{M} \left( \prod_{r=N}^{c+1} \mathbf{G}_{r-1} \big( \theta_{r,c}^{\mathrm Q}, \phi_{r,c}^{\mathrm Q} \big) \right).
\end{equation}
Based on the iteration order in Algorithm~\ref{alg:proposed_arbitrary_revised}, the final working matrix can be expressed as $\mathbf{W}_{\mathrm f}=\mathbf{L}_{\mathrm Q}^\mathrm{H}\mathbf{Q}$. Combining this with $\mathbf{W}_{\mathrm f}=\mathbf{E}_{N,M}\mathbf{D}_{\mathrm Q}$ gives $\label{eq:semi_appendix_conclusion} \mathbf{Q} = \mathbf{L}_{\mathrm Q} \mathbf{E}_{N,M} \mathbf{D}_{\mathrm Q},$ which is exactly \eqref{eq:semi_factorization_revised}. This completes the proof.

\bibliographystyle{IEEEtran}
\bibliography{IEEEabrv,reference.bib}

\begin{thebibliography}{10}
\providecommand{\url}[1]{#1}
\csname url@samestyle\endcsname
\providecommand{\newblock}{\relax}
\providecommand{\bibinfo}[2]{#2}
\providecommand{\BIBentrySTDinterwordspacing}{\spaceskip=0pt\relax}
\providecommand{\BIBentryALTinterwordstretchfactor}{4}
\providecommand{\BIBentryALTinterwordspacing}{\spaceskip=\fontdimen2\font plus
\BIBentryALTinterwordstretchfactor\fontdimen3\font minus
  \fontdimen4\font\relax}
\providecommand{\BIBforeignlanguage}[2]{{%
\expandafter\ifx\csname l@#1\endcsname\relax
\typeout{** WARNING: IEEEtran.bst: No hyphenation pattern has been}%
\typeout{** loaded for the language `#1'. Using the pattern for}%
\typeout{** the default language instead.}%
\else
\language=\csname l@#1\endcsname
\fi
#2}}
\providecommand{\BIBdecl}{\relax}
\BIBdecl

\bibitem{ericssonMobilityReport2025}
\BIBentryALTinterwordspacing
{Ericsson}, ``Ericsson mobility report,'' Ericsson, Tech. Rep., Nov. 2025.
  [Online]. Available:
  \url{https://www.ericsson.com/en/reports-and-papers/mobility-report}
\BIBentrySTDinterwordspacing

\bibitem{iturM2160}
\BIBentryALTinterwordspacing
{ITU-R}, ``Framework and overall objectives of the future development of {IMT}
  for 2030 and beyond,'' International Telecommunication Union, Tech. Rep.
  Recommendation ITU-R M.2160-0, Nov. 2023. [Online]. Available:
  \url{https://www.itu.int/rec/R-REC-M.2160-0-202311-I/en}
\BIBentrySTDinterwordspacing

\bibitem{tataria2021}
H.~Tataria \emph{et~al.}, ``{6G} wireless systems: Vision, requirements,
  challenges, insights, and opportunities,'' \emph{Proceedings of the IEEE},
  vol. 109, no.~7, pp. 1166--1199, Jul. 2021.

\bibitem{letaief2019}
K.~B. Letaief, W.~Chen, Y.~Shi, J.~Zhang, and Y.-J.~A. Zhang, ``The roadmap to
  {6G}: {AI} empowered wireless networks,'' \emph{IEEE Communications
  Magazine}, vol.~57, no.~8, pp. 84--90, Aug. 2019.

\bibitem{butler1961}
J.~Butler and R.~Lowe, ``Beam-forming matrix simplifies design of
  electronically scanned antennas,'' \emph{Electronic Design}, vol.~9, pp.
  170--173, Apr. 1961.

\bibitem{rotman1963}
W.~Rotman and R.~F. Turner, ``Wide-angle microwave lens for line source
  applications,'' \emph{IRE Transactions on Antennas and Propagation}, vol.~11,
  no.~6, pp. 623--632, Nov. 1963.

\bibitem{zeng2014lens}
Y.~Zeng, R.~Zhang, and Z.~N. Chen, ``Electromagnetic lens-focusing antenna
  enabled massive {MIMO}: Performance improvement and cost reduction,''
  \emph{IEEE Journal on Selected Areas in Communications}, vol.~32, no.~6, pp.
  1194--1206, Jun. 2014.

\bibitem{elayach2014spatial}
O.~E. Ayach, S.~Rajagopal, S.~Abu-Surra, Z.~Pi, and R.~W. Heath, ``Spatially
  sparse precoding in millimeter wave {MIMO} systems,'' \emph{IEEE Transactions
  on Wireless Communications}, vol.~13, no.~3, pp. 1499--1513, Mar. 2014.

\bibitem{alkhateeb2015limited}
A.~Alkhateeb, G.~Leus, and R.~W. Heath, ``Limited feedback hybrid precoding for
  multi-user millimeter wave systems,'' \emph{IEEE Transactions on Wireless
  Communications}, vol.~14, no.~11, pp. 6481--6494, Nov. 2015.

\bibitem{gao2016hybrid}
X.~Gao, L.~Dai, S.~Han, C.-L. I, and R.~W. Heath, ``Energy-efficient hybrid
  analog and digital precoding for mmwave {MIMO} systems with large antenna
  arrays,'' \emph{IEEE Journal on Selected Areas in Communications}, vol.~34,
  no.~4, pp. 998--1009, Apr. 2016.

\bibitem{molisch2017hybrid}
A.~F. Molisch \emph{et~al.}, ``Hybrid beamforming for massive {MIMO}: A
  survey,'' \emph{IEEE Communications Magazine}, vol.~55, no.~9, pp. 134--141,
  Sep. 2017.

\bibitem{tang2022pathloss}
W.~Tang \emph{et~al.}, ``Path loss modeling and measurements for reconfigurable
  intelligent surfaces in the millimeter-wave frequency band,'' \emph{IEEE
  Transactions on Communications}, vol.~70, no.~9, pp. 6259--6276, Sep. 2022.

\bibitem{chen2023channel}
W.~Chen, C.-K. Wen, X.~Li, and S.~Jin, ``Channel customization for joint
  {Tx-RISs-Rx} design in hybrid mmwave systems,'' \emph{IEEE Transactions on
  Wireless Communications}, vol.~22, no.~11, pp. 8304--8319, Nov. 2023.

\bibitem{deng2021holographic}
R.~Deng \emph{et~al.}, ``Reconfigurable holographic surfaces for future
  wireless communications,'' \emph{IEEE Wireless Communications}, vol.~28,
  no.~6, pp. 126--131, Dec. 2021.

\bibitem{new2025fas}
W.~K. New \emph{et~al.}, ``A tutorial on fluid antenna system for {6G}
  networks: Encompassing communication theory, optimization methods and
  hardware designs,'' \emph{IEEE Communications Surveys \& Tutorials}, vol.~27,
  no.~4, pp. 2325--2377, 2025.

\bibitem{wong2026pixel}
K.-K. Wong, C.~Wang, S.~Shen, C.-B. Chae, and R.~Murch, ``Reconfigurable pixel
  antennas meet fluid antenna systems: A paradigm shift to electromagnetic
  signal and information processing,'' \emph{IEEE Wireless Communications},
  vol.~33, no.~1, pp. 191--198, Feb. 2026.

\bibitem{zhu2024movable}
L.~Zhu, W.~Ma, and R.~Zhang, ``Movable antennas for wireless communication:
  Opportunities and challenges,'' \emph{IEEE Communications Magazine}, vol.~62,
  no.~6, pp. 114--120, Jun. 2024.

\bibitem{shao2025sixdma}
X.~Shao, Q.~Jiang, and R.~Zhang, ``{6D} movable antenna based on user
  distribution: Modeling and optimization,'' \emph{IEEE Transactions on
  Wireless Communications}, vol.~24, no.~1, pp. 355--370, Jan. 2025.

\bibitem{castellanos2026trihybrid}
M.~R. Castellanos, S.~Yang, C.-B. Chae, and R.~W. Heath, ``Embracing
  reconfigurable antennas in the tri-hybrid {MIMO} architecture for {6G} and
  beyond,'' \emph{IEEE Transactions on Communications}, vol.~74, pp. 381--401,
  2026.

\bibitem{heath2026trihybrid}
R.~W. Heath \emph{et~al.}, ``The tri-hybrid {MIMO} architecture,'' \emph{IEEE
  Wireless Communications}, vol.~33, no.~1, pp. 199--206, 2026.

\bibitem{dardari2026ota}
D.~Dardari, G.~Torcolacci, G.~Pasolini, and N.~Decarli, ``Over-the-air
  electromagnetic signal processing: From fundamentals to emerging paradigms,''
  \emph{IEEE Signal Processing Magazine}, vol.~43, no.~1, pp. 6--28, Jan. 2026.

\bibitem{lin2018diffractive}
X.~Lin \emph{et~al.}, ``All-optical machine learning using diffractive deep
  neural networks,'' \emph{Science}, vol. 361, no. 6406, pp. 1004--1008, Sep.
  2018.

\bibitem{an2023jsac}
J.~An \emph{et~al.}, ``Stacked intelligent metasurfaces for efficient
  holographic {MIMO} communications in {6G},'' \emph{IEEE Journal on Selected
  Areas in Communications}, vol.~41, no.~8, pp. 2380--2396, Aug. 2023.

\bibitem{an2025twc}
J.~An, M.~D. Renzo, M.~Debbah, H.~V. Poor, and C.~Yuen, ``Stacked intelligent
  metasurfaces for multiuser downlink beamforming in the wave domain,''
  \emph{IEEE Transactions on Wireless Communications}, vol.~24, no.~7, pp.
  5525--5538, Jul. 2025.

\bibitem{gu2024pdnn}
Z.~Gu, Q.~Ma, X.~Gao, J.~W. You, and T.~J. Cui, ``Direct electromagnetic
  information processing with planar diffractive neural network,''
  \emph{Science Advances}, vol.~10, no.~29, p. eado3937, Jul. 2024.

\bibitem{teng2025pdnn}
X.~Teng \emph{et~al.}, ``Planar diffractive neural networks empowered
  communications: A spatial modulation scheme,'' arXiv preprint
  arXiv:2512.00847, 2025.

\bibitem{nerini2025analog}
M.~Nerini and B.~Clerckx, ``Analog computing for signal processing and
  communications---part {I}: Computing with microwave networks,'' \emph{IEEE
  Transactions on Signal Processing}, vol.~73, pp. 5183--5197, 2025.

\bibitem{nerini2025analog2}
------, ``Analog computing for signal processing and communications---part
  {II}: Toward gigantic {MIMO} beamforming,'' \emph{IEEE Transactions on Signal
  Processing}, vol.~73, pp. 5198--5212, 2025.

\bibitem{keshavarz2025programmable}
R.~Keshavarz, K.~Zelaya, N.~Shariati, and M.-A. Miri, ``Programmable circuits
  for analog matrix computations,'' \emph{Nature Communications}, vol.~16, p.
  8514, 2025.

\bibitem{reck1994}
M.~Reck, A.~Zeilinger, H.~J. Bernstein, and P.~Bertani, ``Experimental
  realization of any discrete unitary operator,'' \emph{Physical Review
  Letters}, vol.~73, no.~1, pp. 58--61, Jul. 1994.

\bibitem{clements2016}
W.~R. Clements, P.~C. Humphreys, B.~J. Metcalf, W.~S. Kolthammer, and I.~A.
  Walmsley, ``Optimal design for universal multiport interferometers,''
  \emph{Optica}, vol.~3, no.~12, pp. 1460--1465, Dec. 2016.

\bibitem{miller2012}
D.~A.~B. Miller, ``All linear optical devices are mode converters,''
  \emph{Optics Express}, vol.~20, no.~21, pp. 23\,985--23\,993, Oct. 2012.

\bibitem{shen2017nanophotonic}
Y.~Shen \emph{et~al.}, ``Deep learning with coherent nanophotonic circuits,''
  \emph{Nature Photonics}, vol.~11, no.~7, pp. 441--446, Jul. 2017.

\bibitem{golub2013matrix}
G.~H. Golub and C.~F.~V. Loan, \emph{Matrix Computations}, 4th~ed.\hskip 1em
  plus 0.5em minus 0.4em\relax Baltimore, MD, USA: Johns Hopkins University
  Press, 2013.

\bibitem{liang2014hybrid}
L.~Liang, W.~Xu, and X.~Dong, ``Low-complexity hybrid precoding in massive
  multiuser {MIMO} systems,'' \emph{IEEE Wireless Communications Letters},
  vol.~3, no.~6, pp. 653--656, Dec. 2014.

\end{thebibliography}

\end{document}